\documentclass[11pt]{article}

\usepackage{graphicx}
\usepackage{epstopdf}
\usepackage{amsmath}
\usepackage{amssymb}
\usepackage{amsfonts}
\usepackage{amsthm}
\usepackage[usenames]{color}
\usepackage{array}

\usepackage[
      colorlinks=true,
      linkcolor=blue,
      urlcolor=blue,
      filecolor=blue,
      citecolor=red,
      pdfstartview=FitV,
      pdftitle={},
      pdfauthor={},
      pdfsubject={},
      pdfkeywords={},
      pdfpagemode=None,
      bookmarksopen=true
]{hyperref}

\usepackage{epsfig}

\usepackage{hyperref}

\def\beq{\begin{eqnarray}}
\def\eeq{\end{eqnarray}}

\def \RR{{\mathbb{R}}}

\def\mf{\mathfrak}
\def\be{\begin{equation}}
\def\ee{\end{equation}}
\def\bea{\begin{eqnarray}}
\def\eea{\end{eqnarray}}

\def\be{\begin{equation}}
\def\ee{\end{equation}}
\def\bea{\begin{eqnarray}}
\def\eea{\end{eqnarray}}

\newcommand{\rom}[1]{\mathrm{#1}}

\def\cJ{\mathcal{J}}

\def\cL{\mathcal{L}}
\def\cM{\mathcal{M}}

\def\cP{\mathcal{P}}
\def\cQ{\mathcal{Q}}

\def\cV{\mathcal{V}}

\def\f{\frac}

\def\mf{\mathfrak}
\def\nn{\nonumber}

\numberwithin{equation}{section}

\begin{document}

\begin{flushright}
\texttt{\today}
\end{flushright}

\begin{centering}

\vspace{2cm}

\textbf{\Large{
Non-supersymmetric Microstates of the MSW System}}

 \vspace{0.8cm}

  {\large  Souvik Banerjee$^a$, Borun D. Chowdhury$^b$, \\
  Bert Vercnocke$^c$, Amitabh Virmani$^d$}

  \vspace{0.5cm}

\begin{minipage}{.9\textwidth}\small
\begin{center}

{\it $^a$ Centre for Theoretical Physics, University of Groningen,\\
Nijenborgh 4, The Netherlands.}\\
{\it $^b$ Department of Physics,
Arizona State University \\
Tempe, Arizona 85287, USA
}

{\it $^c$ Institute of Physics, University of Amsterdam,\\
Science Park, Postbus  94485, 1090 GL Amsterdam, Netherlands}\\
{\it $^d$ Institute of Physics \\
Sachivalaya Marg, Bhubaneswar, Odisha, India 751005} \\
  \vspace{0.5cm}
{\tt souvik.banerjee@rug.nl,  bdchowdh@asu.edu, bert.vercnocke@uva.nl, virmani@iopb.res.in}
\\ $ \, $ \\

\end{center}
\end{minipage}

%\end{center}

\begin{abstract}
We present an analysis parallel to that of Giusto, Ross, and Saxena (arXiv:0708.3845) and construct a discrete family of non-supersymmetric microstate geometries of the Maldacena-Strominger-Witten system. The supergravity configuration in which we look for the smooth microstates is constructed using $\mathrm{SO}(4,4)$ dualities applied to an appropriate seed solution. The $\mathrm{SO}(4,4)$ approach offers certain technical advantages. Our microstate solutions are smooth in five dimensions, as opposed to all previously known non-supersymmetric microstates with AdS$_3$ cores, which are smooth only in six dimensions. The decoupled geometries for our microstates are related to global AdS$_3 \times \mathrm{S}^2$ by spectral flows.
\end{abstract}

\end{centering}

\newpage

\tableofcontents

\newpage

\section{Introduction}
 Although string theory has made significant progress on counting the number of microstates responsible for  black hole entropy~\cite{Sen:2014aja}, it has not yet given a satisfactory answer about their gravitational description. An outstanding open problem in the physics of black holes is to understand the gravitational description of such microstates. In the last decade or so, developments inspired by AdS/CFT and supergravity have offered us tools to explore these issues. The best studied example in this regard is the supersymmetric D1-D5 system.

Studies related to the D1-D5 system broadly fall into two parts: (i) those related to the five-dimensional black hole with 2-charges, and (ii) those related to the five-dimensional black hole with 3-charges.  The gravitational description of the microstates of the 2-charge black hole has been explicitly constructed~\cite{Balasubramanian:2000rt,
Maldacena:2000dr,
Lunin:2001fv,
Lunin:2001jy,
Lunin:2002bj,
Taylor:2005db,
Skenderis:2006ah,
Kanitscheider:2007wq}
and their entropy has been successfully accounted for~\cite{Palmer:2004gu,Rychkov:2005ji} (see also \cite{Bak:2004rj, Bak:2004kz}).
However, since the horizon area of the corresponding black hole in the absence of higher-derivative corrections to supergravity is exactly zero, the true nature of the black hole entropy requires further exploration. See for instance references~\cite{Giusto:2004xm, Sen:2009bm} for discussions on this point. For the 3-charge black hole in five-dimensions, many explicit microstates are known
\cite{
Mathur:2003hj,
Giusto:2004id,
Giusto:2004ip,
Giusto:2004kj,
Bena:2005va,
Berglund:2005vb,
Bena:2006is}. The gravitational description of these so-called `microstate geometries'---smooth horizonless solutions in supergravity---is possible due to non-trivial topology and charges dissolved in magnetic fluxes, as recently highlighted in~\cite{Gibbons:2013tqa}. However, a detailed understanding of the space of solutions and the validity of the supergravity approximation in describing such states is still far from complete. Similar developments have been made for other systems as well.
This collection of ideas go under the name of the ``fuzzball'' proposal~\cite{Mathur:2005zp}. We refer the reader to reviews
\cite{
Mathur:2005zp,
Mathur:2005ai,
Balasubramanian:2006gi,
Bena:2007kg,
Chowdhury:2010ct,
Bena:2013dka}
for detailed discussions and further references on these ideas.

The results mentioned above are all for supersymmetric black holes. It is an important and daunting task to extend the success of supersymmetric solutions to more general non-supersymmetric settings, in particular to non-extremal black holes. These are qualitatively different from supersymmetric solutions as now the black holes have non-zero temperature, so they Hawking radiate. Only a handful of solutions are known that can potentially describe appropriate non-supersymmetric microstates~\cite{Jejjala:2005yu, Giusto:2007tt, AlAlawi:2009qe, Bena:2009qv}.\footnote{There is a considerable body of work on extremal, non-supersymmetric smooth and non-smooth multi-center solutions, see e.g.,~\cite{
Goldstein:2008fq,
Bena:2009en,
Bena:2009ev,
DallAgata:2010dy,
Bossard:2011kz}, and reference~\cite{DallAgata:2011nh} for a review.
We do not consider solutions of these type in this paper.}
Among these the solutions of references~\cite{Jejjala:2005yu,Giusto:2007tt,AlAlawi:2009qe} have near core AdS$_3 \times \mathrm{S}^3$ regions. These geometries are smooth in six dimensions. A natural problem is to extend such studies to four-dimensional black holes (or five dimensional black strings). In the BPS setting some such analysis already exist~\cite{deBoer:2008fk}. A construction of non-supersymmetric microstates that are smooth in five dimensions has not been attempted.

In this paper we aim to find non-supersymmetric smooth solutions with core region as AdS$_3 \times \mathrm{S}^2$.  Our interest in AdS$_3 \times \mathrm{S}^2$ arises from the fact that M-theory on AdS$_3 \times \mathrm{S}^2 \times \mathrm{T}^6$ is dual to a 1+1 dimensional Maldacena-Strominger-Witten (MSW) CFT~\cite{Maldacena:1997de}. MSW CFT is qualitatively different from the D1-D5 CFT. The previous studies of non-supersymmetric smooth solutions~\cite{Jejjala:2005yu,Giusto:2007tt,AlAlawi:2009qe}
have focused on the D1-D5 system. Our results establish the existence of non-supersymmetric microstate geometries in five dimensions that correspond to states in the MSW CFT.
We expect that by adding  further charges on our microstate geometries it should be possible to construct dual of MSW states that can potentially describe microstates of certain 4d black hole.

The MSW CFT has (4,0) supersymmetry and is much less understood than the D1-D5 CFT. We hope that through supergravity analysis like the one performed in this paper one can make some (at least qualitative) progress on understanding the nature of the MSW CFT. We base this expectation on the fruitful application of a bottom-up approach to the D1-D5 system, in particular by considering microstate geometries. It was very instructive to consider the Maldacena-Maoz~\cite{Maldacena:2000dr} solution and its various generalizations
\cite{
Mathur:2003hj,
Giusto:2004id,
Giusto:2004ip,
Giusto:2004kj,
Jejjala:2005yu}. Such studies improved our understanding of short and long string sectors, various twist sectors, and the orbifold nature of the D1-D5 CFT. Furthermore, the study of the non-BPS solutions of~\cite{Jejjala:2005yu} led to the identification of their classical instability~\cite{Cardoso:2005gj} with Bose-enhanced Hawking radiation~\cite{Chowdhury:2007jx}. A clear understanding of the similar issues for the MSW system is still lacking in the literature.

Unfortunately, there is no systematic way of going about finding new five-dimensional non-supersymmetric smooth solutions.
We follow the same approach as~\cite{Jejjala:2005yu, Giusto:2007tt}, where we first construct a suitable general family of supergravity configurations, and then find constraints on the parameters to obtain smooth solutions.
As far as technicalities of our construction are concerned, our set-up is directly related to the set-up studied by Giusto, Ross, and Saxena (GRS)~\cite{Giusto:2007tt}. GRS study non-supersymmetric microstates of the D1-D5-KK system, which upon dimensional reduction over a circle and dualities is related to our set-up. Many details of our analysis are however different from theirs. In particular, details related to the study of the BPS limit, smoothness analysis, and decoupling limit differ. In addition we explicitly show that the decoupled geometries for our microstates are related to global AdS$_3 \times \mathrm{S}^2$ by spectral flows and we also compute the corresponding conformal weights in the MSW CFT. The analog of this study in the context of D1-D5-KK CFT  is missing  in~\cite{Giusto:2007tt}.

The rest of the paper is organised as follows. In section \ref{summary} we present a concise and self-contained summary of the supergravity configuration we work with. A detailed construction of these fields is relegated to appendices \ref{SugraConstruction} and \ref{GroupTheory}. In section \ref{BPSlimit1} we discuss the BPS limit of the configuration presented in section \ref{summary} and discuss its smoothness.  In section \ref{BPSlimit2} we show how to obtain the (singular) M5-M5-M5 MSW black string limit. This M5-M5-M5 limit is precisely the ``reverse process'' of what Bena and Warner~\cite{Bena:2005va, Bena:2007kg} have called ``geometric transition."  In section \ref{sec:find_smooth} via a systematic search in the parameter space we find values of parameters at which all singularities can be removed. These solutions are specified by the three M5 charges, one M2 charge, and an integer parameter $n$. For $n=1$ the solution reduces to the supersymmetric solution discussed in section \ref{BPSlimit}. For all values $n > 1$ the solutions are non-supersymmetric. In this section we also perform a detailed regularity analysis, and show that this discrete family of solutions is free from curvature singularities, horizons, and closed time-like curves. We also show that all the matter fields are regular and that the
metric is regular at what GRS call ``corners.''  In section \ref{sec:decouple} we take a decoupling limit of the solution and exhibit AdS$_3 \times \mathrm{S}^2$ as the near core geometry. In this section we also show that the decoupled geometries are related to global AdS$_3 \times \mathrm{S}^2$ by spectral flows, and compute the corresponding conformal weights in the MSW CFT. We end in section \ref{disc} with a summary and a brief discussion of open problems.

\section{Supergravity configuration}
\label{summary}
In this section we write the five-dimensional solution that will be the basis for the smooth microstates in the later parts of the paper. It is a solution of five-dimensional U(1)$^3$ supergravity, the so-called STU truncation of the maximal five-dimensional supergravity where one keeps only two vector multiplets. This section is self-contained --- we explicitly list all notation required to write the metric and the matter fields. Details of the construction of this configuration, including the uplift to IIB theory on a T$^4$, can be found in appendices \ref{SugraConstruction} and \ref{GroupTheory}.

We have in coordinates $(\rho, x = \cos \theta, \phi, t, z_5)$
\bea
 ds^2_5 &=&   f^2(dz_5 + \check A^0_{[1]})^2+  f^{-1} ds_4^2, \\
ds_4^2 &=& -
e^{2 U} (dt + \omega_3)^2 + e^{-2 U}
ds_3^2, \\
ds^2_3 &=&  \frac{F}{\Delta} d \rho^2 + \frac{F}{1-x^2} dx^2 + (1-x^2) \Delta d\phi^2,
\eea
where
\begin{align}
 f & = \frac{\sqrt{AD}}{ (\tilde H_2 \tilde H_3 (D+ A s_2^2 s_3^2
V_0^2))^{1/3}}, &
e^{2 U} &= \frac{F}{\sqrt{AD}}, \\
F &= \rho^2 + m^2 - b^2 x^2, &  \Delta &= \rho^2 + m^2 - b^2.
\end{align}
The various quantities appearing in these expressions are as follows
\bea
A &=& F + 2 p \left[ \rho + \frac{(pq -m^2)}{(p + q)} + b
\frac{\sqrt{p^2 + m^2}\sqrt{q^2 + m^2}}{m (p + q)} x\right], \\
B &=& F + 2 q \left[ \rho + \frac{(pq -m^2)}{(p + q)} - b
\frac{\sqrt{p^2 + m^2}\sqrt{q^2 + m^2}}{m (p + q)} x\right], \\
C &=& 2 \frac{\sqrt{q}}{\sqrt{p + q}} \left[ \sqrt{q^2 + m^2} (\rho + p)
- q \frac{\sqrt{p^2 + m^2}}{m} b x \right],\\
D &=& B c_2^2 c_3^2 -F (c_2^2 s_3^2 + s_2^2 c_3^2) + \frac{F^2}{B}
s_2^2 s_3^2 - \frac{C^2 F}{A B}  s_2^2 s_3^2, \\
V_0 &=& -\frac{1}{A}\sqrt{\frac{q (q^2 + m^2)}{p (p^2 + m^2)}}\left[ F + 2 p \left( \rho + p + \frac{q b}{m} \sqrt{\frac{p^2 + m^2}{q^2 + m^2}} x\right) \right],\\
\tilde H_{2,3} &=& A + (A - G) s_{2,3}^2,\\
G &=& \frac{A F -C^2}{B},
\eea
where
\begin{align}
s_{2,3} &= \sinh \alpha_{2,3}, &  c_{2,3} &= \cosh \alpha_{2,3}.
\end{align}
The one-form $\check A^0_{[1]}$ is
\be
\check A^0_{[1]} = \zeta_0 (dt  + \omega_3)  +  A^0_3,
\ee
\begin{align}
\zeta_0 &= c_2 c_3 s_2 s_3 \frac{C V_0}{D}, & A^0_3 &= s_2 s_3 \kappa^1_0, &
\omega_3 &= c_2 c_3 \omega^0,
\end{align}
where the three-dimensional one-forms are
\bea
\kappa^1_0 &=& \frac{2 b \sqrt{q} \sqrt{p+q}  \left(1- x^2 \right)}{m
 \sqrt{m^2+p^2} F} \left((p-q+\rho ) m^2+p q \rho
 \right) d\phi, \\
\omega^0 &=& -\frac{2 b \sqrt{p} \sqrt{q} \left(1 - x^2\right) \left(m^2
(p+q+\rho )-p q \rho \right)}{m (p+q) F} d \phi.
\eea
The five-dimensional vectors are
\begin{equation}
A^I_{[1]} = \chi^I(dz+ \check A^0_{[1]})+ \check A^I_{[1]} ,
\end{equation}
with
\begin{align}
\chi_1 &= - \frac{A s_2 s_3 V_0}{D + A s_2^2 s_3^2 V_0^2}, &
\chi_2 &=  - c_2 s_3 \frac{C}{\tilde H_3}, &
\chi_3 &=  - c_3 s_2 \frac{C}{\tilde H_2},
 \end{align}
and the one-forms $\check A^I_{[1]}$ are
\begin{equation}
\check A^I_{[1]} = \zeta_I(dt+\omega_3) + A_3^I,
\end{equation}
with
\begin{align}
\zeta_1 &= c_2 c_3 \frac{C}{D}, &
\zeta_2 &= s_2 c_3 (B c_2^2 - F s_2^2) \frac{V_0}{D},&
\zeta_3 &= s_3 c_2 (B c_3^2 - F s_3^2) \frac{V_0}{D},
 \end{align}
and
\begin{align}
A^1_3    &= \omega^1, &
A^2_3 &= -s_2 c_2 \kappa^{0}_{0}, &
A^3_3 &= -s_3 c_3 \kappa^{0}_{0}.
\end{align}
The remaining three-dimensional one-forms are
\bea
\omega^1 &=& \frac{2 \sqrt{p}}{\sqrt{p+q}
   F} \left[\sqrt{m^2+p^2}  \Delta x- \frac{b\sqrt{m^2+q^2}}{m}
   (1-x^2) (p \rho - m^2)\right] d \phi,  \\
\kappa^{0}_{0} &=& - \frac{2}{F} q  \left[ \Delta x +
\sqrt{\frac{q^2 + m^2}{p^2 + m^2}} \frac{b}{m} (p \rho - m^2)(1-x^2)\right]
d\phi.
\eea
Finally, the five-dimensional U(1)$^3$ supergravity scalars are $h^I = y^I/f$ with
\begin{align}
y_1 &= \frac{\sqrt{AD}}{D + A s_2^2 s_3^2 V_0^2}, &
y_2 &= \frac{\sqrt{A D}}{\tilde H_3}, &
y_3 &= \frac{\sqrt{A D}}{\tilde H_2}.
 \end{align}

The uplift of this rather complicated solution to six dimensions directly matches the configuration in~\cite{Giusto:2007tt}. The above solution is asymptotically flat in four dimensions. Physical quantities in four dimensions such as mass, angular momentum, electric and magnetic charges are (with four-dimensional Newton's constant take to be unity)
\bea
\cM &=& \frac{1}{2} \left[ p + q ( 1+ s_2^2 + s_3^2) \right], \\
\cJ &=& b \frac{\sqrt{pq}(p q - m^2)}{m (p + q)}c_2 c_3, \\
\cQ &=& \sqrt{\frac{q (q^2 + m^2)}{p + q}} c_2 c_3, \\
\cP &=& \sqrt{\frac{p (p^2 + m^2)}{p + q}},\\
\cQ_2 &=& q s_2 c_2, \\
\cQ_3 &=& q s_3 c_3,
\eea
where we use the notation $\cP, \cQ_2, \cQ_3$ to denote the 3 magnetic (M5) charges and $\cQ$ to denote the electric (M2) charge.

\section{BPS limit}
\label{BPSlimit}

In order to make contact with the previous literature related to the above configuration, let us consider the uplift to  six dimensions and take the limit $m \to 0$ with various charges and angular momentum held fixed.
This BPS limit allows us to connect rather directly to the five-dimensional analysis in the following.

In coordinates
\begin{align}
\tilde r &= \rho - b x, & \tilde x &= \frac{\rho x - b}{\rho - b x},
\end{align}
the metric in the BPS limit in the Bena-Warner~\cite{Bena:2007kg} form is\footnote{We have fixed a typo in the 6d BPS metric compared to~\cite{Giusto:2007tt}. The term $-(dt + k)^2$ in the metric is incorrectly
typed in~\cite{Giusto:2007tt} as $-(dt - k)^2$.}:
\bea
ds^2_6{}_\rom{BPS} &=& (Z_2 Z_3)^{-1/2} \left[ - (dt + k)^2 + (dz_5 + \omega_P - k)^2 \right] + (Z_2 Z_3)^{1/2} ds_\rom{B}^2,  \\
ds_\rom{B}^2 &=&  V^{-1}(d z_6+ \vec \chi)^2 + V \left(d\tilde r^2 + \tilde r^2 \frac{d \tilde x^2}{1-\tilde x^2} + \tilde r^2 (1 - \tilde x^2) d\phi^2 \right).
\eea
Here $Z_2, Z_3,$ and $V$ are harmonic functions on the three-dimensional flat base space spanned by $\tilde r, \tilde x, \phi$ coordinates; $k$ and $\omega_P$ are one-forms on
the four-dimensional Gibbons-Hawking base space spanned by  $z_6, \tilde r, \tilde x, \phi$. From the IIB supergravity point of view, which is obtained by adding T$^4$ to the six-dimensional uplift,
$\cP, \cQ, \cQ_2, \cQ_3$ respectively denote the BPS limits of KK monopole ($z_6$ as KK fibre), KK electric ($z_6$ direction), D1 ($z_5$ direction), and D5
($z_5$ and four-torus  directions)
 charges.

The expressions for the 1-forms are
\begin{align}
& k = \left(H_K + \frac{H_P}{2V}\right) (d z_6 + \vec \chi) + \vec k, & & \star_3  d \vec k = V dH_K - H_K dV - \frac{1}{2} dH_P,\label{eq:dk}
\\
&\omega_P = \frac{H_P}{V}(d z_6 + \vec \chi) + \vec \omega_P, & & \star_3  d \vec \omega_P =  - d H_P,\label{eq:domegaP}
\end{align}
and the harmonic functions are\footnote{Upon KK reduction on $z_5$, the fields can be matched to a M2-M2-KKM system. In this frame, the Bena-Warner harmonic functions are
\begin{align}
L_1 &= 1, & L_2 &= Z_2, & L_3 &= Z_3, \\
K^1 &= H_P, &  K^2 & = K^3 = 0, & M &= H_K,
\end{align}
and the function $V$ is same as in \eqref{functionsVZ}.}
\begin{align}
& V = 1 + \frac{2 \cP}{\tilde r}, &  & Z_i  = 1 + \frac{2 \cQ_i}{\Sigma}, \quad i = 2, 3, \label{functionsVZ}\\
& H_K = - \frac{\cQ}{2 \cP} \left(1 + \frac{2\cP}{\Sigma} \right), & & H_P = \frac{\cQ}{\cP} + \frac{2 \cQ_2 \cQ_3}{\cQ} \left(\frac{1}{\tilde r} - \frac{1}{\Sigma}\right), \label{functionHP}
\end{align}
with
\begin{align}
& \vec  \chi =  2 \cP x d\phi, & & \Sigma = \sqrt{\tilde r^2 + 4 b^2 + 4 b \tilde r \tilde x}.
\end{align}
Integrations of \eqref{eq:dk} and \eqref{eq:domegaP} give
\begin{align}
&  \vec \omega_P  =  \left[-\frac{2 \cQ_2 \cQ_3}{\cQ} x + \frac{2 \cQ_2 \cQ_3}{\cQ} \frac{1}{r_c}(r x + 2 b) \right] d \phi,  \\
&  \vec k =  \left[- \frac{\cP \cQ^2 + b (\cQ_2 \cQ_3 -\cQ^2 ) x}{b \cQ}  + \frac{\cP \cQ^2 (r + 2 b x) -b (\cQ^2 - \cQ_2 \cQ_3) (2 b + r x) }{b \cQ r_c}\right] d\phi,
\end{align}
and integrations of $\star_3 d \vec Z_i = d Z_i$ give
\be
 \vec Z_i =  \frac{2 \cQ_i}{r_c} (r x + 2 b) d\phi.
\ee

A detailed regularity analysis of the above BPS solution is presented in~\cite{Bena:2005ay},
and is concisely summarized in~\cite{Giusto:2007tt}. We do not reproduce this analysis here.

\subsection{Five dimensions}
\label{BPSlimit1}
We reduce the above BPS solution to five-dimensions by performing KK reduction over $z_6$.
This new five-dimensional solution
can also be written in the Bena-Warner~\cite{Bena:2007kg} form. It simply corresponds to the BPS limit of configuration of section \ref{summary}. We introduce a new set of coordinates
\begin{align}
t &= - v & z_5 &= -u + v. \label{uvcoordinates}
\end{align}
In these coordinates the metric is
\be
ds^2_5 = - f^2 (dv + k)^2 + f^{-1}
\left[
H_P^{-1} \left(du - \vec \omega_P\right)^2
+ H_P \left(d\tilde r^2 + \tilde r^2 \frac{d \tilde x^2}{1-\tilde x^2} + \tilde r^2 (1-\tilde x^2) d\phi^2\right)\right],
\ee
with $\star_3 d \vec \omega_P = - d H_P$. The one-form $k$ is
\be
k = \left[\frac{V (Z_2 Z_3 - H_K H_P)}{H_P^2} - \frac{1}{2} \right] (du -  \vec \omega_P) - \vec k d\phi,
\ee
and
\be
f = \left[ \frac{H_P^3}{V^2 Z_2 Z_3 (Z_2 Z_3 - 2 H_K H_P)}\right]^\frac{1}{3}.
\ee
The above metric can be realized in the Bena-Warner notation in an M5-M5-M5 frame with an M2 charge as
\begin{align}
& K^1{}_\rom{BW}   = V &
& K^{2}{}_\rom{BW}  = Z_{2} &
& K^{3}{}_\rom{BW}  = Z_{3} &
& V{}_\rom{BW}  =  H_P &\\
& L_{1,\rom{BW}}  = - 2 H_K &
& L_{2,\rom{BW}}  = 0 &
& L_{3,\rom{BW}}  = 0 &
& M{}_\rom{BW} =  - \frac{1}{2}.
\end{align}
We use the subscript ``BW'' to avoid confusion with the various other frames used above.

The five-dimensional vectors in this Bena-Warner presentation are
\be
A^{I} =  - \frac{dt + k}{Z_{I, \rom{BW}}} + \frac{K^I_\rom{BW}}{V_\rom{BW}} (du - \omega_P) + \vec \xi^{(I)}
\ee
where $\star_3  d\vec \xi^{(I)}  = - d K^{I}{}_\rom{BW}$, which can be integrated to
\begin{align}
&\xi^{(1)} = - 2 \cP x d \phi, & & \xi^{(2)} = - \vec Z_2, & & \xi^{(3)} = - \vec Z_3, &
\end{align}
and where
\be
Z_{I, \rom{BW}} = \frac{1}{2} |\epsilon_{IJK}| \frac{K^J_\rom{BW}K^K_\rom{BW}}{V_{\rom{BW}}} + L_{I,\rom{BW}}.
\ee
The scalars are simply
\be
h^I = \frac{(Z_{1, \rom{BW}}Z_{2, \rom{BW}}Z_{3, \rom{BW}})^\frac{1}{3}}{Z_{I, \rom{BW}}}.
\ee
Note in particular that $V_\rom{BW} = H_P$ and hence the Gibbons-Hawking base for the five-dimensional BPS solution has two centers, see equation \eqref{functionHP}.
\paragraph{Regularity analysis:} A regularity analysis proceeds along the same lines as for the 6d metric as in~\cite{Bena:2005ay} and~\cite{Giusto:2007tt}.
\begin{itemize}
 \item For the smoothness of the 5d metric we must ensure that $k$ is a smooth 1-form on the base space:  it must
vanish at $\tilde r= 0$ and $\Sigma =0$ where the harmonic function
$V_\rom{BW}$ diverges. Requiring this (more precisely $k_u=0$) fixes the distance between the two-centers to be
\be
b = \frac{2 \cP \cQ^2}{\cQ_2 \cQ_3 - \cQ^2}. \label{distance}
\ee

\item The requirement that the $u$-circle smoothly shrinks to zero size at $\tilde r = 0$ and at $\Sigma = 0$ fixes the identification $u \sim u + 2 \pi\Delta u$ and hence also the radius $R_{u}$ of the $u$-circle at infinity
\be
\Delta u = \frac{4 \cQ_2 \cQ_3 }{\cQ}\,,\qquad R_{u} = \frac{4 \cQ_2 \cQ_3 \cP^\frac{2}{3}}{\cQ(\cP^2 + \cQ^2)^{\frac{1}{3}}}.
\ee

This concludes the BPS limit of the five-dimensional solution.

\end{itemize}
\subsection{M5-M5-M5 limit}
\label{BPSlimit2}
Now let us consider taking the limit where the two centres become coincident, while keeping the 3 M5 charges fixed. In this limit  we see from \eqref{distance}  that the M2 charge disappears. Thus, we expect to recover a (singular) MSW black
string with only M5 charges. We now examine how this limit is achieved.

We start by noting that the Bena-Warner class of solutions have the following symmetry:
\begin{align}
& t' = \lambda t, & & \psi' = \psi /\lambda, & &  V'  = V/\lambda, \\
& L_I'  = \lambda L_I & & K^I{}'  = K^I, & & M'  = \lambda^2 M.
\end{align}
Under these rescalings the one-form $k$, $\vec \chi$, etc.~scale appropriately so that the full metric remains invariant.

We rescale the Bena-Warner smooth solution with $\lambda = \frac{\cQ}{\cP}$ and then take the limit $\cQ \to 0$.
The full five-dimensional metric  in the rescaled coordinates becomes
\bea
ds^2 &=& - (K^1K^2K^3)^{-4/3} (dt' + k)^2 + (K^1K^2K^3)^{2/3} ds^2_\rom{base} \\
k &=& (K^1K^2K^3) d\psi',
\eea
where the base space metric takes the form
\be
ds^2_\rom{base} = d \psi'^2 + (ds^2_3)_\rom{flat},
\ee
Writing in terms of the original coordinates we get
\be
ds^2 = - 2 (K^1K^2K^3)^{-1/3} dt d \psi  + (K^1K^2K^3)^{2/3}(ds^2_3)_\rom{flat},
\ee
which is precisely the metric of the M5-M5-M5 black string in the double null coordinates.

The reader may wonder how a limit of a smooth solution gives rise to a (singular) black string. The answer lies in the fact that the limiting black string is of zero-entropy. This is precisely the ``reverse process'' of what Bena and Warner~\cite{Bena:2005va, Bena:2007kg} have called ``geometric transition."  In their set-up a zero entropy black string  goes into a 2-center smooth solution via nucleation of Gibbons-Hawking centres. In our set-up when we take the limit of centres to be coincident we recover the singular black string metric.

\section{Smooth solutions and regularity analysis}
\label{sec:find_smooth}
Now we want to explore whether there are any smooth solutions in the family of five-dimensional metrics we have constructed. We follow the same reasoning as~\cite{Giusto:2007tt}.
By inspection we can see that the metric has potential singularities at
$\tilde H_2 =0, \tilde H_3 =0, A =0, D=0, \Delta = 0, F=0, B=0, (D + A s_2^2 s_3^2 V_0^2) = 0$, and $x = \pm 1$. We will focus on the singularity at $\Delta = 0$ and will try to interpret it as a smooth origin. The singularities at $x = \pm 1$ are like the usual coordinate singularities arising from the degeneration of azimuthal coordinate at the poles of the two-sphere. The analysis of the $x = \pm 1$ singularities will require appropriate identifications, which will be discussed later. We expect other singularities to be true curvature singularities, so we will arrange parameters in the smooth solutions such that  $\tilde H_2, \tilde H_3, A, D, F, B > 0$ everywhere. Moreover we require the identifications of $u$ and $\phi$ to lie in a surface of constant $v$ (see equation \eqref{uvcoordinates}).

It turns out that in performing the analysis outlined in the previous paragraph, there are no changes compared to what is already done in~\cite{Giusto:2007tt}. We will see that our analysis starts to differ from the next subsection. For brevity we do not write all equations again, and present only the most relevant details. Interpreting $\Delta = 0$ at $\rho =  \rho_0$ as a smooth origin,
 where a spatial direction goes to zero size, we need
\begin{equation}
\rho_0 = \sqrt{b^2-m^2} = \frac{m^2 (p+q)}{p q - m^2},
\end{equation}
which implies
\begin{equation}
 b^2=\frac{m^2 \left(m^2+p^2\right) \left(m^2+q^2\right)}{\left(p q-m^2\right)^2}. \label{bvalue}
\end{equation}
We will always work with the positive square-root for \eqref{bvalue}. It can be checked that no other singularities are encountered in the region $\rho > \rho_0, -1 \le x \le 1$ as  $\tilde H_2, \tilde H_3, A, D, F, B$ are all strictly positive\footnote{Showing that $D$ is positive is quite non-trivial. In~\cite{Giusto:2007tt} it is shown that $D > 0$ by an indirect argument by first showing that the geometry has no closed timelike curves. See also section \ref{sec:CTCs} below.}.

Therefore, we conclude that when constraint \eqref{bvalue} is satisfied, the only singularities in the metric are at $(i)$ $\rho = \rho_0$, $(ii)$ $x =+1$, and $(iii)$  $x =-1$. Each of these singularities is a degeneration of a spatial cycle in the $(u, \phi)$ part of the metric. We now turn to an analysis of these singularities.

\subsection{Global identifications}
\label{sec:find_smooth_global}
We want to find the global identifications in $(u,\phi)$ space that yield a smooth metric. We investigate the three singularities one by one.  We write the general Killing vector in the $(u,\phi)$ space as
\be
\xi =\partial_{\phi}-\alpha \partial_u,
\ee
where the choice
 \be
 \alpha = - s_2 s_3 \kappa_{0,\phi}^1,
 \ee
ensures the vanishing of the norm $\xi \cdot \xi$ of the Killing vector at the following one-cycle degeneracies:

\paragraph{Degeneracy I ($x = +1$):}
For $x = +1$, $F=\Delta$, $\omega^{0}=0$ and $\kappa_{0,\phi}^1 = 0$. Hence $\alpha = 0$.
We note that locally the relevant part of the metric looks like (up to an over-all constant)
\be
\frac{dx^2}{1-x^2} + (1-x^2) d\phi^2,
\ee that says that $\phi$ has to have a period of $2\pi$.
Hence the identification required for obtaining a smooth origin is at $x = +1$
\begin{equation}
(u,\phi) \sim (u, \phi + 2\pi).
\label{phi-iden}
\end{equation}

\paragraph{Degeneracy II ($x = -1$):}
For $x = -1$ the discussion is identical to the $x = +1$ case. Again, $F=\Delta$, $\omega^{0}=0$ and $\kappa_{0,\phi}^1 = 0$ and hence $\alpha = 0$.
The metric is locally flat yielding the identification of $\phi$ to be \eqref{phi-iden}.

\paragraph{Degeneracy III ($\rho = \rho_0$):}
At $\rho = \rho_0$, $F=b^2(1-x^2)$, $\omega^{0}=0$ and $\kappa_{0,\phi}^1 = \frac{4 q \sqrt{q}\sqrt{p +q}}{\sqrt{q^2+m^2}}$. Hence, $\alpha = -4 s_2 s_3 \frac{q \sqrt{q}\sqrt{p+q}}{\sqrt{q^2+m^2}}$. Therefore, the relevant circle is along $\phi$ at fixed $u - 4 s_2 s_3\frac{ q \sqrt{q}\sqrt{p+q}}{\sqrt{q^2+m^2}} \phi$.
The leading contribution (of order $(\rho-\rho_0)$) to the non-zero size of the circle away from $\rho = \rho_0$ comes only from the term $\frac{\Delta}{F}(1-x^2) d\phi^2$ in the metric. Defining a new coordinate $Z$ as $\rho =\rho_0 (1 + 2 Z^2)$, the relevant part of the metric becomes
 \begin{equation}
 \frac{4}{\rho_0} (\tilde H_2 \tilde H_3 (D+A s_2^2 s_3^2 V_0^2))^{\frac{1}{3}}
 \left(dZ^2 + Z^2 \frac{\rho_0^2}{b^2} d\phi^2 \right).
 \end{equation}
From this we note that $\phi$ must be periodic with periodicity $2 \pi n$ where $n = \frac{b}{\rho_0}.$
Hence, the identification in this case is:
\be
(u, \phi) \sim (u + 8 \pi n s_2 s_3 q \frac{\sqrt{q}\sqrt{p+q}}{{\sqrt{q^2+m^2}}}, \phi + 2 n \pi).
\label{u-iden}
 \ee

Upon KK reduction over $u$ we expect to obtain a four-dimensional asymptotically flat geometry, so $\phi$ must have periodicity $2 \pi$. Hence given the identification
\eqref{u-iden} we obtain another constraint on the parameters (in addition to \eqref{bvalue}) that
\begin{equation}
n = \frac{b}{\rho_0} \label{integer}
\end{equation}
must be an integer. Thus, our smooth solutions are characterized by five parameters, which we can take to be the three magnetic charges, one electric charge and the integer $n$. This counting works as follows: in our supergravity configurations we have six parameters $(p, q, m, b, \alpha_2, \alpha_3)$ and we have one constraint \eqref{bvalue} and one relation \eqref{integer}.

\subsection{Regularity analysis}

In this section we present a detailed regularity analysis of the smooth solutions found in the previous section.

\subsubsection{Corners}
From the analysis presented in section \ref{sec:find_smooth_global}, it follows that the discrete family of solutions is smooth at $\rho = \rho_0$ or  at $x=\pm 1$.\footnote{Recall that throughout this paper we are using the coordinate $x$ in place of the azimuthal coordinate $\theta$ on the two-sphere. The two are simply related by $x= \cos \theta$.}  However, from that analysis it is not clear what happens at the `corners'~\cite{Giusto:2007tt}, where  $\rho = \rho_0$ and $x=\pm 1$. In this section we  introduce explicit coordinates and show that the eleven-dimensional geometry is smooth near these points as well.

\paragraph{Corner 1:} Let us consider the first corner $\rho = \rho_0$, $x = + 1$. We start by defining new coordinates
\bea
\tilde r &=& (\rho - \rho_0) + \rho_0 (1-x), \\
\tilde x &=& (1+x) \left[ \frac{\rho - \rho_0}{\tilde r}\right] - 1,
\eea
with the property that $\rho = \rho_0$, $x = + 1$ is at $\tilde r = 0$, with $\rho = \rho_0$, $x \neq 1$ along $\tilde x = -1$, and $\rho \neq \rho_0$, $x = +1$  along $\tilde x = +1$. The inverse relation for these coordinates is
\bea
\rho &=& \frac{1}{2} \left( \sqrt{\tilde r^2 + 4 \tilde r \tilde x \rho_0 + 4\rho_0^2} +  \tilde r\right),\\
x &=& \frac{1}{2 \rho_0} \left(  \sqrt{\tilde r^2 + 4 \tilde r \tilde x \rho_0 + 4\rho_0^2} - \tilde r \right),
\eea
which is of direct use for the computations below. Near $\tilde r = 0$, the factor $\frac{f e^{2U}}{F}$ goes as a constant. For this reason we consider the following scaled metric
\bea
\frac{f e^{2U}}{F} ds_5^2 &=& \frac{f^3 e^{2U}}{F}\left( - du + dv + s_2 s_3 \kappa_0^1 d \phi + c_2 c_3 s_2 s_3 \frac{C V_0}{D} (-dv + c_2 c_3 \omega^0 d\phi)\right)^2 \nonumber \\
&&
+ \frac{dx^2}{1-x^2} + \frac{d\rho^2}{\Delta} + (1-x^2)\frac{\Delta}{F} d\phi^2 - \frac{e^{4U}}{F} (-dv + c_2 c_3 \omega^0 d\phi)^2.
\label{scaled_metric}
\eea
In the limit $\tilde r \to 0$ various pieces in the last two terms of the above metric have the following limiting behavior
\bea
\frac{e^{4U}}{F} &\approx& \mbox{constant} \\
\omega_0 &\approx& \mathcal{O}(\tilde r) \\
(1-x^2)\frac{\Delta}{F} &=& \frac{(1-\tilde x^2)\tilde r}{(1+ \tilde x+ n^2(1-\tilde x))\rho_0} + \mathcal{O}(\tilde r^2).
\eea
From these expressions we see that (i) the size of the $v$ direction remains finite as we zoom near the first corner, (ii) all contributions from the $\omega_0$ terms can be ignored, as they are subleading.

Now let us examine the first line of the scaled metric \eqref{scaled_metric}. After some computation we note that
\bea
\frac{f^3 e^{2U}}{F} &=& \frac{(1 + \tilde x + n^2 (1-\tilde x)) (p + \rho_0 - n^2 \rho_0)^2}{16 (n^2 -1)^2 s_2^2 s_3^2 \rho_0^2 (p + \rho_0)^3} \tilde r + \mathcal{O}(\tilde r^2) \\
\frac{C V_0}{D} &\approx & \mathcal{O}(1).
\eea
Thus we see that from the first line of the scaled metric \eqref{scaled_metric} only the following piece is relevant near the corner
\be
\frac{f^3 e^{2U}}{F}\left(-du + s_2 s_3 \kappa_0^1 d \phi \right)^2.
\ee
The limiting value of $\kappa_0^1$ near the corner is
\be
\kappa_0^1 = \frac{4 n \left(n^2-1\right) (1-\tilde x) (p+\rho_0) \sqrt{\rho_0 (p+\rho_0)}}{\left(1+\tilde x + n^2 (1-\tilde x)\right) \left(p+\rho_0 -n^2\rho_0\right)}+\mathcal{O}\left(\tilde r\right),
\ee
Now a straightforward (if somewhat tedious) computation shows that
\bea
&&\frac{f^3 e^{2U}}{F}\left(-du + s_2 s_3 \kappa_0^1 d \phi \right)^2  + \frac{dx^2}{1-x^2} + \frac{d\rho^2}{\Delta} + (1-x^2)\frac{\Delta}{F} d\phi^2 \approx \nn \\
&&
\frac{1}{2 \rho_0 \tilde r} \Big{[} d \tilde r^2 + \tilde r^2 \frac{d \tilde x^2}{1-\tilde x^2}
 + \frac{\tilde r^2}{2 n^2} (1 + \tilde x) \left( d\phi + \frac{d \hat u}{2 \cP} \right)^2  + \frac{\tilde r^2}{2} (1 - \tilde x) \left( d\phi - \frac{d \hat u}{2 \cP} \right)^2\Big{]},
\label{R4metric}
\eea
where the new coordinate $\hat u$ is
\be
 \hat u = \frac{1}{d_1} u + \frac{d_2}{d_1} \phi,
\ee
with the constants $d_1$ and $d_2$ given as
\bea
d_1 &=&p s_2 s_3  \frac{(n^2 -1)(p + \rho_0)\sqrt{\rho_0 (p + \rho_0)}}{n (p (p + \rho_0 - n^2 \rho_0))^{3/2}}, \\
d_2 &=& -2 p s_2 s_3\frac{(n^2 -1) (p + \rho_0)\sqrt{\rho_0 (p + \rho_0)}}{n (p (p + \rho_0 - n^2 \rho_0))}.
\eea
The identification \eqref{u-iden} becomes
\be
(\hat u, \phi) \sim (4 \pi n\cP, 2 \pi n).
\label{u-iden2}
\ee
Now, if we define $\tilde r = R^2$, $\tilde x = 2 \hat x^2 -1$, and
\be
\psi_1 = \frac{1}{2} \left( \phi - \frac{\hat u}{2 \cP} \right), \qquad \qquad \psi_2 = \frac{1}{2n} \left(\phi + \frac{\hat u}{2 \cP} \right),
\ee
the metric \eqref{R4metric} becomes the standard metric on $\mathbb{R}^4$, while the identifications \eqref{phi-iden} and \eqref{u-iden2} become respectively $\psi_1 \sim  \psi_1 + 2 \pi$, $\psi_2 \sim \psi_2 + 2 \pi$. Thus the local geometry near the first corner is globally $\mathbb{R}^4$, and hence smooth. The five-dimensional scalars of the U(1)$^3$ supergravity have finite non-zero limiting values, hence the eleven dimensional geometry is also smooth near the corner.

\paragraph{Corner 2:} Now let us consider the second corner $\rho = \rho_0$, $x = - 1$. Again we start by defining new coordinates
\bea
\tilde r &=& (\rho - \rho_0) + \rho_0 (1+x), \\
\tilde x &=& (1-x) \left[ \frac{\rho - \rho_0}{\tilde r}\right] - 1.
\eea
In these coordinates $\rho = \rho_0$, $x = - 1$ is at $\tilde r = 0$, with $\rho = \rho_0$, $x \neq -1$ along $\tilde x = -1$, and $\rho \neq \rho_0$, $x = -1$  along $\tilde x = +1$. The inverse relation for these coordinates is
\bea
\rho &=& \frac{1}{2} \left( \sqrt{\tilde r^2 + 4 \tilde r \tilde x \rho_0 + 4\rho_0^2} +  \tilde r\right),\\
x &=& - \frac{1}{2 \rho_0} \left(  \sqrt{\tilde r^2 + 4 \tilde r \tilde x \rho_0 + 4\rho_0^2} - \tilde r \right),
\eea
which we can use below. Near $\tilde r = 0$, the factor $\frac{f e^{2U}}{F}$ again goes as a constant. For this reason we again consider the scaled metric \eqref{scaled_metric}. In the limit $\tilde r \to 0$ various pieces in the last two terms of the scaled metric have the following limiting behavior
\bea
\frac{e^{4U}}{F} &\approx& \mbox{constant} \\
\omega_0 &\approx& \mathcal{O}(\tilde r) \\
(1-x^2)\frac{\Delta}{F} &=& \frac{(1-\tilde x^2)\tilde r}{(1+ \tilde x+ n^2(1-\tilde x))\rho_0} + \mathcal{O}(\tilde r^2).
\eea
From these expressions we see that (i) the size of the $v$ direction remains finite as we zoom near the first corner, (ii) all contributions from the $\omega_0$ terms can be ignored, as they give subleading contributions.

The rest of the analysis proceeds  exactly the same way as for the first corner.  We conclude that the local geometry near the second corner is globally $\mathbb{R}^4$, and hence smooth. The five-dimensional scalars of the U(1)$^3$ supergravity also have finite non-zero limiting values, hence the eleven dimensional geometry is also smooth near the second corner.
\subsubsection{Closed timelike curves}
\label{sec:CTCs}
In the $(u, v)$ coordinates introduced in equation \eqref{uvcoordinates}
we verify the absence of closed timelike curves for the five-dimensional metric. We do this by showing that $v$ is a global time function, which requires $g^{vv} < 0$ everywhere. The $vv$ component of the five-dimensional inverse metric is
\be
g^{vv} = - \frac{f}{e^{2 U}} + \frac{f e^{2 U}}{\Delta (1-x^2)} c_2^2 c_3^2 (\omega_\phi^0)^2.
\ee
To show that this quantity is negative, we rewrite it as (after substituting the values for $e^{2 U}$ and $f$)
\be
g^{vv} = \frac{1}{(\tilde H_2 \tilde H_3 (D+ A s_2^2 s_3^2
V_0^2))^{1/3}} \left[  - \frac{AD}{F} + \frac{F}{\Delta (1-x^2)} c_2^2 c_3^2 (\omega_\phi^0)^2 \right].
\label{gvv}
\ee

From the analysis of section 5.3 of~\cite{Giusto:2007tt} it follows that the quantity in the square brackets is negative. Since that analysis applies as it is to our case, we do not present any details on this, and refer the reader to that reference. We note two minor typos in section 5.3 of~\cite{Giusto:2007tt} \begin{enumerate}
\item
A factor of 4 missing in their equation (5.47): term involving $J^2$ should read, $-4 B J^2 (\rho-\rho_0)^2 \sin^2\theta$.
\item In equation (5.65) tilde on $c_2$ should not be there; the left-hand side  should be $c_2(-1)$.
\end{enumerate}

Now it remains to show that the overall prefactor in equation \eqref{gvv} is positive. From previous sections we already know that $\tilde H_2$, $\tilde H_3$, $F$, $\Delta$, and $A$ are all positive for $\rho > \rho_0$. Since the quantity in the square parenthesis in equation \eqref{gvv} is negative, it follows that $D > 0$. Hence it follows that the prefactor in equation \eqref{gvv} is positive.

\subsubsection{Vectors and scalars}

The scalars $h^I$ are regular and positive everywhere, including at the `corners'  where $\rho=\rho_0$ and $x = \pm 1$.

The vectors of the solution have potential singularities where various cycles degenerate.  To analyse this let us note that at the different degeneration points the various one-forms appearing in the solution are
\begin{align}
x &= +1:         &\omega^0_\phi &= 0,  &\omega^1_\phi &= +2 {\cal P}, &\kappa^0_{0,\phi} &= -2 q, &\kappa^1_{0,\phi} &= 0, \\
x &= -1:         &\omega^0_\phi &= 0,  &\omega^1_\phi &= -2 {\cal P}, &\kappa^0_{0,\phi} &= +2 q, &\kappa^1_{0,\phi} &= 0, \\
\rho &= \rho_0:  &\omega^0_\phi &= 0,  &\omega^1_\phi &= -2 {\cal P}, &\kappa^0_{0,\phi} &= -2 q, &\kappa^1_{0,\phi} &= 4 q \frac{ \sqrt{q}\sqrt{p+q}}{\sqrt{q^2+m^2}}.
\end{align}

The gauge fields along the degenerating direction are regular only if the integrals around the degenerating cycles can be made  zero by a (possibly) large gauge transformation. These integrations are
\begin{align}
\oint_\phi (\xi \cdot A^I)|_{x=+1}        &= 2\pi(+2{\cal P},+2 s_2 c_2 q, +2 s_3 c_3 q),
\label{eq:integral_cycles1} \\
\oint_\phi (\xi \cdot A^I)|_{x=-1}        &= 2\pi(-2{\cal P},-2 s_2 c_2 q, -2 s_3 c_3 q),
\label{eq:integral_cycles2} \\
\oint_\phi (\xi \cdot A^I)|_{\rho=\rho_0} &= 2\pi(-2{\cal P},+2 s_2 c_2 q, +2 s_3 c_3 q).
\label{eq:integral_cycles3}
\end{align}
These integrals simply denote non-trivial Wilson lines in the limits when the respective cycles degenerate. The non-zero values of these integrals can all be removed by appropriate \emph{large} gauge transformations; they represent the usual behaviour of the gauge field at the location of a magnetic monopole. Moreover, the first of these equations \eqref{eq:integral_cycles1} can be related to the second \eqref{eq:integral_cycles2} by a large gauge transformation consistent with the Dirac quantization of magnetic charges on the two-sphere spanned by $(x,\phi)$.

In this sense all matter fields are regular (or can be made regular) by appropriate large gauge transformations.

\section{Decoupling limit}
\label{sec:decouple}
We already know from~\cite{Giusto:2007tt} that the six-dimensional lift of the solution that we have constructed qualitatively looks like a smooth D1-D5 solution sitting at the tip of the Kaluza-Klein monopole. This explains why the decoupling limit in six-dimensions gives an AdS$_3 \times $ S$^3$ geometry, as shown in~\cite{Giusto:2007tt}.
Upon dualities the `six-dimensional' D1-D5-KKM system can be related to the `five-dimensional' M5-M5-M5 system. Therefore, from the five-dimensional solution studied in this paper, we expect a decoupling limit  where we obtain an AdS$_3 \times $ S$^2$ geometry. In this section we exhibit this.

As in the previous studies~\cite{Jejjala:2005yu, Giusto:2007tt} obtaining a decoupling limit  requires us to scale some charges in a way such that we go close to extremality.
More precisely, one is required to take the so-called dilute gas limit from the point of view of the dual CFT.  This is achieved by taking $\rho_0 \to 0$, while keeping the three magnetic charges fixed. As we take this limit, we also  change coordinates so as to zoom in on the core region by setting
\bea
\rho &=& \rho_0 r, \\
u &=& \frac{2\chi}{d \sqrt{\rho_0}}, \label{uscale}\\
v &=& \frac{\tau}{d \sqrt{\rho_0}}, \label{vscale}
\eea
and keeping $r, \chi$, and $\tau$ fixed.  In equations \eqref{uscale} and \eqref{vscale} $d$ is a constant
\be
d = \frac{1}{2\sqrt{\cP \cQ_2 \cQ_3}}.
\ee
In this limit the metric becomes an AdS$_3 \times $ S$^2$ geometry. This can be made explicit by introducing further new coordinates
\bea
r &=& 1 + 2 R^2, \\
x &=& \cos \theta,\\
\tilde \tau &=& \frac{\tau}{2},\\
\varphi &=& \chi - \frac{\tau}{2},
\eea
in terms of which the near core metric becomes
\be
ds^2 = \ell_\rom{AdS}^2 \left(- (1+ R^2) d\tilde \tau^2 + \frac{dR^2}{1+ R^2} + R^2 d\varphi^2\right)
+ \ell_\rom{S^2}^2 \left( d\theta^2 + \sin^2 \theta (d\phi + n (d\tilde \tau - d\varphi))^2\right).
\ee
where
\be
\ell_\rom{AdS} = 4 (\cP \cQ_2 \cQ_3)^{1/3} , \qquad \qquad \ell_\rom{S^2} = 2 (\cP \cQ_2 \cQ_3)^{1/3}.
\ee

The scalars are constant in this limit. They take their `attractor values'
\be
h^1 = \frac{\cP}{(\cP \cQ_2 \cQ_3)^{1/3}}, \qquad h^2 = \frac{\cQ_2}{(\cP \cQ_2 \cQ_3)^{1/3}},\qquad h^3 = \frac{\cQ_3}{(\cP \cQ_2 \cQ_3)^{1/3}},
\ee
and the vectors in the decoupling limit are
\bea
A^1_\rom{decoupling} &=& - 2 n \cP d \chi + 2 \cP x (d\phi + n d \tilde \tau - n d \varphi), \\
A^2_\rom{decoupling} &=& + 2 n \cQ_2 d \chi + 2 \cQ_2 x (d\phi + n d \tilde \tau - n d \varphi), \\
A^3_\rom{decoupling} &=& + 2  n \cQ_3 d \chi + 2 \cQ_3 x (d\phi + n d \tilde \tau - n d \varphi).
\eea
The identifications \eqref{phi-iden} and \eqref{u-iden} respectively become
\bea
(\varphi, \phi) &\sim& (\varphi, \phi + 2 \pi), \\
(\varphi, \phi) &\sim& (\varphi + 2 \pi, \phi + 2 n \pi).
\eea
Hence geometry of the decoupled solution is global AdS$_3$.

The above expressions fit into the (generic) asymptotic behaviour of fields analysed in~\cite{deBoer:2008fk}. Our decoupled geometry is obtained by taking a decoupling limit of a non-BPS solution. For this reason the CFT analysis of~\cite{deBoer:2008fk}, which is done for BPS states, cannot be taken over directly. Instead we should view our states as excitations above those studied in~\cite{deBoer:2008fk}. From the above AdS$_3 \times$ S$^2$ geometry we can read off the CFT quantum numbers of the dual state. In the following we present such an analysis.

Let us start by recalling the integer quantization of M2 and M5 charges. If we normalize the charges in the five-dimensional KK setting (asymptotically $\mathbb{R}^{3,1} \times S^1$) as
\be
P^I = \frac{1}{2\pi}\int_{S^2} F^I, \qquad \qquad Q_I = \frac{1}{16 \pi G_5} \int_{S^2 \times S^1} (h^I)^{-2} \star_5 F^I,
\ee
where the integrals are calculated at spatial infinity, then the integer quantizations are~\cite{Emparan:2006mm}
\be
n^I = \left(\frac{\pi}{4 G_5}\right)^\frac{1}{3} P^I,  \qquad \qquad N_I = \left(\frac{4 G_5}{\pi}\right)^\frac{1}{3} Q_I.
\ee
Note that the central charge is
\be
c  = \frac{3 \ell_\rom{AdS}}{2 G_3} = 6 n^1 n^2 n^3 = D_{IJK}n^I n^J n^K,
\ee
where $D_{IJK} = |\epsilon_{IJK}|$. Furthermore we find
\be
N_1 = n n^2 n^3, \qquad \qquad  N_2 = 0,\qquad \qquad N_3 = 0.
\ee

We can now use the results from~\cite{deBoer:2008fk} to compute contributions to the left and right sector conformal weights from the gravitational sector, the U(1) charges, and the SU(2) sector. Recall that the SU(2) sector arises upon dimensional reduction to AdS$_3$, and corresponds to angular momentum in four dimensions.

The gravitational contributions comes just from the Casimir energy of the cylinder dual to global AdS$_3$
\bea
L_0^\rom{grav} = \bar L_0^\rom{grav} = -\f{n^1 n^2 n^3}{4}.
\eea
The SU(2) charge contributes only to the left sector and is given in terms of the SU(2) current
\be
L_0^{\mathrm{SU}(2)}= \f{J^2}{4 n^1 n^2 n^3}.
\ee
The U(1) currents are proportional to the M2 charges. We have
\be
J_I = N_I.
\ee
They contribute to the energy in both the left and the right sector. To write this succinctly it is convenient to define
\be
(D^+)^{IJ} = \f{n^I n^J}{12 n^1 n^2 n^3}, \qquad (D^-)^{IJ}=D^{IJ} - (D^+)^{IJ}.
\ee
where $D^{IJ}= (D_{IJK} n^K)^{-1}$. Then the contributions to the conformal weights are
\bea
L_0^{\mathrm{U}(1)}= (D^+)^{IJ} J_I J_J, \\
\bar L_0^{\mathrm{U}(1)}= -(D^-)^{IJ} J_I J_J.
\eea
The total weights are simply the sum of these three contributions\footnote{There is a curious feature that $L_0 + \bar L_0 = \f{1}{6} n^2 n^1 n^2 n^3 + \Delta M$, where $\Delta M$ is the ADM energy obtained by subtracting the contribution from the M5 branes: $\Delta M= M_\rom{ADM} - M_\rom{M5}.$ This is unlike the solutions in~\cite{Jejjala:2005yu} where the CFT energy is equal to the corresponding $\Delta M$. \label{footnote}}
\bea
L_0 &=& -\f{n^1 n^2 n^3}{4} + (D^+)^{IJ} J_I J_J + \f{J^2}{4 n^1 n^2 n^3}, \\
&=&-\f{n^1 n^2 n^3}{4}  + \f{(n)^2 n^1 n^2 n^3}{3},\\
\bar L_0 &=& -\f{n^1 n^2 n^3}{4} -(D^-)^{IJ} J_I J_J, \\
&=&-\f{n^1 n^2 n^3}{4}  + \f{(n)^2 n^1 n^2 n^3}{3}.
\eea

The symmetry algebra underlying the MSW CFT admits an SU(2) spectral flow
\bea
J \to J+\eta, \qquad & L_0 \to L_0 + \f{1}{2 n^1 n^2 n^3} \eta J + \f{1}{4 n^1 n^2 n^3} \eta^2,
\eea
and three U(1) spectral flows~\cite{deBoer:2006vg, deBoer:2008fk}
\bea
J_I \to J_I + \eta_I, \qquad & L_0 \to L_0 + 2 (D^+)^{IJ} \eta_I J_J + (D^+)^{IJ} \eta_I \eta_J, \\
& \bar L_0 \to \bar L_0 - 2 (D^-)^{IJ} \eta_I J_J - (D^-)^{IJ} \eta_I \eta_J.
\eea

For our solution
 the only non-trivial currents are
\be
J_1 = n n^2n^3, \qquad \qquad J = n n^1 n^2 n^3.
\ee
Therefore, our decoupled solution is related to global AdS$_3 \times \mathrm{S}^2$ by spectral
flows through parameters
\be
\eta_1 = n n^2 n^3, \qquad \qquad \eta  = n n^1n^2n^3.
\ee
From this we also conclude that for our decoupled solution the spectral flow invariant weights receive contributions only from the gravitational sector.

\section{Discussion}
\label{disc}
In this paper we have constructed a discrete family of non-supersymmetric microstates of the MSW system. It is quite non-trivial that we are able to find such smooth microstates. Our microstates are smooth in five-dimensions and have AdS$_3$ cores. This is different from all previously known non-supersymmetric microstates, which are smooth only in six dimensions.

There are several papers that study smooth solutions for the $(4,0)$ D1-D5-KK system. Since this system is closely related to the MSW system, in table \ref{table1} we attempt to give an overview of this literature. We only focus on solutions that have near core AdS$_3$ regions, and classify them by the amount of spectral flows the near core region exhibits. Given the relation of this system to the MSW system via string dualities, we expect a general such solution to have four integer parameters in the STU model: one for the left SU(2) spectral flow and \emph{three} for the U(1) spectral flows. It is a challenging exercise to explicitly construct this configuration\footnote{In the MSW CFT the SU(2) spectral flow acts on the left sector but the U(1) spectral flows act on both the left and the right sectors~\cite{Kraus:2006wn,deBoer:2008fk}.}.
\begin{table}[!h]
\begin{center}
\begin{tabular}{|l|c|c|}
\hline
Study & SU(2) spectral flow & U(1) spectral flow  \\ \hline \hline
Bena and Kraus~\cite{Bena:2005ay} & $1$ & $1$ \\ \hline
Saxena, Potvin, Giusto, and Peet~\cite{Saxena:2005uk} & $1$  & $n$ \\ \hline
Giusto, Ross, and Saxena~\cite{Giusto:2007tt} & $n$ & $n$\\ \hline
Ross and Al-Alawi~\cite{AlAlawi:2009qe}& $n$ & $m, m \neq n$\\ \hline
\end{tabular}
\end{center}
\caption{An overview of the literature studying microstates of the $(4,0)$ D1-D5-KK system.}
\label{table1}
\end{table}

There are several ways in which our study can be extended. The first and perhaps the most interesting such extension would be to obtain a more detailed microscopic understanding of the above discussed microstates, in particular to understand why $\Delta M$ and $L_0 + \bar L_0$ do not match, cf footnote \ref{footnote}. It will also be useful to understand which subclass of our geometries admit the appropriate spin structure so that they can truly be thought of as excited states in the Ramond sector of the CFT. We expect a non-trivial answer to this question because naively we expect that only for odd values of the integer $n$ the dual CFT would be in the Ramond sector. Our bosonic supergravity analysis does not capture this physics.

One of  the most interesting properties
of the JMaRT solution~\cite{Jejjala:2005yu}
is that it has an ergoregion. Horizonless geometries with an ergoregion are excepted to be generically unstable~\cite{Friedman:1978jk}. Therefore it is clearly of interest to study the presence of  ergoregions for our solution. For the six-dimensional version of the solution such a study is complicated~\cite{Giusto:2007tt}. For the five-dimensional solution studied in this paper it is more manageable. However, using intuition from the results for the JMaRT solution~\cite{Chowdhury:2007jx}, we expect spontaneous emission of M2 branes to be the most relevant for our set-up. To study spontaneous emission of M2 branes one effectively needs to study the ergoregion in six dimensions, which  remains an important open problem.

A different five-dimensional as well as other dimensionally reduced versions of the JMaRT solution were studied in \cite{Gimon:2007ps}. It was found that the five-dimensional version of the solution is smooth except at two points where there are non-supersymmetric orbifold singularities. Their study is similar to ours with one notable difference that in their study, dimensional reduction from 6d to 5d is done over $z_5$ as opposed to $z_6$.\footnote{We thank Simon Ross for pointing this out to us.} Extrapolating the results of \cite{Gimon:2007ps} to the six-dimensional GRS configuration, we do not expect to obtain a smooth solution upon dimensional reduction over $z_5$. We expect the presence of non-supersymmetric orbifold singularities in the spacetime. On the other hand, in this paper we have shown that upon dimensional reduction over $z_6$ we do obtain a smooth solution. If we do further dimensional reductions, say to four or to three dimensions, we anticipate that the details of the resulting geometries will be similar to the ones studied in \cite{Gimon:2007ps}.

Finally, it is clearly of interest to construct more general smooth solutions. The simplest such extensions are to add more M2 charges to the configuration considered in this paper Using the SO(4,4) approach used in this paper this line of investigation is clearly within reach. However, to construct qualitatively different configurations, say, one that becomes in the BPS limit a three-centre Bena-Warner solution requires novel techniques. In this regard the inverse scattering method recently developed for  STU supergravity~\cite{Katsimpouri:2012ky, Katsimpouri:2013wka} is the most promising.  We hope to report on some of these issues in our future work.

\subsubsection*{Acknowledgements}
We have benefited from discussions with Iosif Bena, Stefano Giusto, Samir Mathur, Simon Ross, Ashoke Sen and Yogesh Srivastava. AV would like to thank in particular Simon Ross for many patient explanations on reference~\cite{Giusto:2007tt}. AV would also like to thank AEI Golm and IMSc Chennai for their warm hospitality where part of this work was done. BV acknowledges support from the European Union through the Marie Curie Intra-European fellowship 328652--QM--sing and from the ERC Starting Independent Researcher Grant 240210--String--QCD--BH during the initial stages of this work. BV thanks the University of Groningen for warm hospitality during the final stages of this work. SB thanks IOP Bhubaneswar and University of Amsterdam for their warm hospitality where part of this work was done.

\appendix

\section{Construction of supergravity configuration}
\label{SugraConstruction}
In this appendix we present a construction of the supergravity configuration in which we look for smooth microstates discussed in the main text. The construction
below parallels that of Giusto, Ross, and Saxena (GRS)~\cite{Giusto:2007tt} in its overall logic, however the details are quite different. The key difference is that we use the three-dimensional dualities of the dimensionally reduced STU supergravity, whereas GRS use ten-dimensional S and T dualities. In particular, our specific approach is to use
$\mathrm{SO}(4,4)$ dualities applied to an appropriate ``seed'' solution.
The $\mathrm{SO}(4,4)$ approach offers a number of technical advantages. First, in our approach all dualities are implemented as group rotations, as opposed to the approach of GRS  where only some dualities are implemented as group rotations and the rest as boosts and S- and T-dualities in 10 dimensions. Second, to implement S- and T-dualities in the GRS approach certain complicated Hodge dualizations are required at intermediate steps, the analogs of which are more systematically implemented in our approach.

\subsection{Formalism briefly}
Here we introduce our notation and present the set up we work with. For a more detailed discussion of these things in the same notation as below, we refer the reader to~\cite{Virmani:2012kw}.
 We reduce to three dimensions in steps, giving special attention to five and six dimensions.

Recall that a well known truncation of IIB supergravity on a T$^4$ is
\bea
ds^2_{10, \, \rom{string}} = ds^2_6 + e^{\frac{\Phi}{\sqrt{2}}}
ds_4^2, \qquad
\Phi_{10} = \frac{\Phi}{\sqrt{2}}, \qquad  C^{\rom{RR}}_{[2]} = C_{[2]},
\eea
where $ds^2_4$ is the flat metric on T$^4$ and $C^{\rom{RR}}_{[2]}$ is the Ramond-Ramond two-form.
The truncated Lagrangian becomes
\be
{\cal L}_{6} = R_6 \star_6 1 - \frac{1}{2}\star_6 d\Phi \wedge d  \Phi -
\frac{1}{2} e^{\sqrt{2} \Phi} \star_6 F_{[3]} \wedge F_{[3]},
\label{6d}
\ee
with the field strength
$F_{[3]} =
dC_{[2]}$. Upon further dimensional
reduction on a circle, the six-dimensional theory (\ref{6d}) reduces to the
U(1)$^3$ supergravity in 5d. This dimensional reduction proceeds as follows
\bea
\label{metric6d}
ds^2_6 &=& e^{-\sqrt{\frac{3}{2}} \Psi} (dz_6 + A_{[1]}^1)^2 +
e^{\frac{1}{\sqrt{6}} \Psi} ds^2_5 \\
F_{[3]} &=& F_{[3]}^{\rom{(5d)}} + dA_{[1]}^2\wedge (dz + A_{[1]}^1)
\eea
with
\be
  F_{[3]}^{\rom{(5d)}} = dC_{[2]}^{\rom{(5d)}} -dA^2_{[1]}\wedge A^1_{[1]}.
\ee
After dualization of the 5d 2-form $C_{[2]}^{\rom{(5d)}}$ into a one-form $A^3_{[1]}$ we get the Lagrangian of the five-dimensional U(1)$^3$
supergravity. Details of this construction can be found in for example appendix A of~\cite{Virmani:2012kw}. It takes the form
\bea
{\cal L}_5 &=& R_5 \star_5 1 - \frac{1}{2} \star_5  d \Phi \wedge  d \Phi -
\frac{1}{2} \star_5 d \Psi \wedge d \Psi \nonumber \\
& & - \frac{1}{2} e^{-2 \sqrt{\frac{2}{3}} \Psi} \star_5 F^1_{[2]} \wedge
  F^1_{[2]} - \frac{1}{2} e^{\sqrt{\frac{2}{3}}\Psi + \sqrt{2} \Phi}  \star_5
  F^2_{[2]} \wedge F^2_{[2]} \nonumber \\
&&  - \frac{1}{2} e^{\sqrt{\frac{2}{3}}\Psi - \sqrt{2} \Phi} \star_5
F^3_{[2]}\wedge F^3_{[2]} + A^3_{[1]} \wedge F^2_{[2]} \wedge F^1_{[2]}.
\label{Lag5dU13}
\eea
 where the U(1)$^3$
supergravity scalars are parameterized as
\be
h^1 = e^{\sqrt{\frac{2}{3}} \Psi}, \quad h^2 = e^{-\sqrt{\frac{1}{6}} \Psi -\sqrt{\frac{1}{2}} \Phi}, \quad h^3 = e^{-\sqrt{\frac{1}{6}} \Psi + \sqrt{\frac{1}{2}} \Phi}.
\label{realspecial}
\ee Evidently $h^1 h^2 h^3 = 1$. A manifestly triality invariant form for the 5d Lagrangian can also be readily written.
The 6d field strength $F_{[3]}$
in terms of the 5d fields introduced above is:
\be
F_{[3]}  = - (h^3)^{-2} \star_5 dA^3_{[1]} + dA^2_{[1]} \wedge (dz_6 +
A^1_{[1]}).
\label{RR3}
\ee

Upon further dimensional reduction to
four dimensions via the ansatz
\bea
ds^2_5 &=& f^{2}(dz + \check A^0_{[1]})^2 + f^{-1} ds^2_4, \\
\label{dimred5d4d_1}
A^I_{[1]} &=& \chi^I(dz+ \check A^0_{[1]})+ \check A^I_{[1]},
\label{dimred5d4d_2}
\eea
we get the ${\cal N} =2$ STU model.
For the STU model the number of vector-multiplets is three and the prepotential is
\be
F(X) = - \frac{X^1 X^2 X^3}{X^0}.\label{preSTU}
\ee
We use the gauge fixing condition $X^0 = 1$. The scalars $\chi^I$ obtained from equation \eqref{dimred5d4d_2} and $h^I$ from equation \eqref{realspecial}
together  form the complex scalars $z^I =
X^I/X^0$ of the STU theory as
\be
z^I  = - \chi^I + i f h^I \equiv x^I + i y^I.
\ee

Further dimensional reduction over time gives an SO(4,4)/(SO(2,2) $\times$
SO(2,2)) coset model. To this end we
 parameterize our 4d metric as
\begin{equation}
ds^2_4 =  -e^{2U}(dt+\omega_3)^2 + e^{-2U}ds_3^2,\label{metric4d3d}
\end{equation}
and 4d vectors as
\begin{equation}
\check A^\Lambda_{[1]} = \zeta^\Lambda(dt+\omega_3) + A_3^\Lambda,\label{vector4d3d}
\end{equation}
where $\omega_3$ and $A_3^\Lambda$ are 1-forms in 3d.  We dualize the 1-forms as~\cite{Gaiotto:2007ag, Bossard:2009we}
\begin{equation}
-d\tilde{\zeta}_{\Lambda}  = e^{2U}(\mbox{Im} N)_{\Lambda \Sigma} \star_3 (d{A_3}^{\Sigma} + \zeta^{\Sigma} d\omega_3)+ (\mbox{Re} N)_{\Lambda \Sigma}d\zeta^{\Sigma}
\label{dual1}
\end{equation}
and
\begin{equation}
-d\sigma = 2 e^{4U} \star_3 d \omega_3 - \zeta^{\Lambda} d \tilde{\zeta}_{\Lambda} + \tilde{\zeta}^{\Lambda} d \zeta_{\Lambda} .
\label{dual2}
\end{equation}
From equations \eqref{dual1} and \eqref{dual2} it follows that $\tilde{\zeta}_{\Lambda}$ and $\sigma$ are pseudo-scalars dual to ${A_3}^{\Lambda}$ and $\omega_3$ respectively.
Putting  together various pieces from above, we see that there are in total sixteen three-dimensional scalars
\be
\varphi^a =
 \{U,z^I,\bar z^I,\zeta^\Lambda,\tilde \zeta_\Lambda,\sigma \}. \ee
 These scalars parameterize an $\mathrm{SO}(4,4)/(\mathrm{SO}(2,2)\times \mathrm{SO}(2,2))$ coset model.  The Lagrangian in three dimensions is
\bea
\cL_3 = R \star_3 \mathbf{1} -\frac{1}{2} G_{ab} \partial \varphi^a \partial
\varphi^b,
\label{metric3dcoset}
\eea
where $G_{ab}$ is the bi-invariant metric of signature $(8,8)$ on the coset  $\mathrm{SO}(4,4)/(\mathrm{SO}(2,2)\times \mathrm{SO}(2,2))$.
The metric $G_{ab}$ is an analytic continuation of the Ferrara-Sabharwal c-map~\cite{Ferrara:1989ik}.  Certain details on the parametrization of $G_{ab}$ are collected in appendix \ref{appSO44}.

To construct the supergravity configuration we are interested in, we use the solution generating technique based on the above
obtained SO(4,4)/(SO(2,2) $\times$
SO(2,2)) coset model.  We start with a six-dimensional seed solution and dimensionally reduce is to three dimensions.
The 4d to 3d reduction is done over the time direction. We then act on the seed scalars with an appropriate SO(4,4)
group element. After this group rotation, we obtain a new set of scalars, which can be uplifted to obtain a new solution of the higher-dimensional theory.

\subsection{Giusto, Ross, and Saxena (GRS) configuration}
In this subsection we present the five-dimensional version of  the Giusto, Ross, and Saxena (GRS) six-dimensional configuration~\cite{Giusto:2007tt}. In section \ref{summary} we listed all
functions appearing in the metric and the matter fields explicitly.
\subsubsection{Seed configuration}
As in section 2 of GRS,  we first obtain an appropriate seed solution. We reproduce the seed using our
SO(4,4) approach. Once the seed is obtained adding charges (following section 3 of GRS) is relatively straightforward in the SO(4,4) formalism,
which we implement in the following.
The simplicity of adding charges is one of the key technical advantages our approach offers.

To obtain the seed solution we will use various inputs from references~\cite{Rasheed:1995zv, Larsen:1999pp} as well. The seed solution carries Kaluza-Klein (KK) electric and magnetic charges. In fact, it is  a simple analytic continuation~\cite{Giusto:2007tt} in
the parameter space of the Rasheed-Larsen ~\cite{Rasheed:1995zv, Larsen:1999pp} solution.

\paragraph{Starting metric and starting scalars:}
The starting metric we use is the same as that of GRS in six dimensions
\bea
ds^2 &=& - dt^2 + dz_5^2 + \frac{F}{\rho^2 - (m - b x)^2} \left(d z_6 - \frac{2 m \Delta(m - b x)}{b F} d \phi \right)^2 \nn \\
 & & + (\rho^2 - (m - b x)^2) \left[\frac{d\rho^2}{\Delta} + \frac{d x^2}{1-x^2} + \frac{\Delta}{F} (1-x^2) d\phi^2 \right],
\eea
where
\be
F = \rho^2 + m^2 - b^2 x^2, \qquad \Delta = \rho^2 + m^2 - b^2.
\ee
This is the metric of a Kerr-Bolt instanton trivially lifted to six dimensions by adding two flat directions: $t$ and $z_5$. For calculational simplicity we use
$x$ instead of the polar coordinate $\theta$, the two are related by $x = \cos \theta$. Furthermore, as the names suggest, we use the coordinate $z_6$ to reduce from 6d to 5d, the coordinate $z_5$ to reduce from 5d to 4d and finally the time coordinate $t$ to reduce from 4d to 3d. The three-dimensional base metric for the starting solution is
\be
d s_3^2 = \frac{F}{\Delta} d \rho^2 + \frac{F}{1-x^2} dx^2 + (1-x^2) \Delta d\phi^2,
\ee
and the non-zero three-dimensional scalars are
\bea
U &=& \frac{1}{4} \log \left[\frac{F}{\rho^2 - (m - b x)^2}\right], \\
\tilde \zeta^1 &=& - \frac{2 m \rho}{\rho^2 - (m - b x)^2}, \\
y_1 &=& \frac{1}{y_2} = \frac{1}{y_3} = \left[\frac{\rho^2 - (m - b x)^2}{F}\right]^\frac{1}{2}.
\eea
Given these scalars we construct the  SO(4,4) valued matrix $\cM$ corresponding to the starting solution. The general construction of this matrix is described in
appendix \ref{appSO44}. For the starting solution
it takes the form
\be
\cM = \left(
\begin{array}{cccccccc}
 \tilde f & 0 & 0 & 0 & 0 & 0 & 0 & g \\
 0 & 1 & 0 & 0 & 0 & 0 & 0 & 0 \\
 0 & 0 & 1 & 0 & 0 & 0 & 0 & 0 \\
 0 & 0 & 0 & \tilde f & -g & 0 & 0 & 0 \\
 0 & 0 & 0 & g & \tilde f-\frac{2 b g x}{\rho } & 0 & 0 & 0 \\
 0 & 0 & 0 & 0 & 0 & 1 & 0 & 0 \\
 0 & 0 & 0 & 0 & 0 & 0 & 1 & 0 \\
 -g & 0 & 0 & 0 & 0 & 0 & 0 & \tilde f-\frac{2 b g x}{\rho }
\end{array}
\right),
\ee
where we have introduced notation
\be
\tilde f = 1-\frac{2 m (m-b x)}{m^2-b^2 x^2+\rho ^2}, \qquad g = \frac{2 m \rho
}{m^2-b^2 x^2+\rho ^2}.
\ee
This matrix is similar to the corresponding SO(3,3) matrix used in~\cite{Larsen:1999pp} and it is also similar to the matrix $\chi$ used in~\cite{Giusto:2007tt}. The formalisms are however different as they are based on different higher dimensional Lagrangians.
\paragraph{Group rotation:}
Motivated by the analysis of~\cite{Rasheed:1995zv, Larsen:1999pp} we act with the following group element on the above matrix $\cM$,
\be
g_G =  \exp (\alpha K_{q_1}) \exp (-\beta K_{p^1})  \exp (- \gamma K_0),
\label{group}
\ee
as
\be
\cM' = g_G^\sharp \cM g_G,
\ee
where $g_G^\sharp$ is simply $g_G^{-1}$ in this case (see appendix \ref{appSO44}). As
the name of these generators suggest $K_{q_1}$ generates one electric charge (an M2 charge from the M-theory perspective) and
$K_{p^1}$ generates one magnetic charge (M5 charge), and $K_0$ generates the 4d Lorentzian NUT charge. The necessity of acting with
$K_0$ lies in the fact that the group
element $$\exp (\alpha K_{q_1}) \exp (-\beta K_{p^1})$$
in addition to generating electric and magnetic charges also generates a 4d NUT charge, which can be cancelled by appropriately
acting with $$\exp (- \gamma K_0).$$ To achieve this cancellation of the NUT charge we use the relation~\cite{Rasheed:1995zv, Giusto:2007tt}
\be
\tanh \alpha = \sinh \beta \tan2 \gamma.
\ee
As a next step, we change variables similar to the ones introduced in~\cite{Larsen:1999pp}
\bea
\tan \gamma &=& \sqrt{\frac{p}{q}} \\
\sqrt{p^2 + q^2 - 2 p q \cosh 2 \beta}  & =& m \sinh 2 \beta.
\eea
The second of these equations is a bit unwieldy. We note that it can also be written in the following (more useful) form
\be
e^{2 \beta} = \frac{(\sqrt{p^2 + m^2}-p ) (\sqrt{q^2 + m^2} + q)}{m^2}.
\ee
Substituting $p$ and $q$ in place of $\beta$ and $\gamma$ in the matrix $\cM$ is a somewhat tedious exercise. In the end we obtain an 8 $\times$ 8 matrix from which we can read various 3d scalars.

\paragraph{Three dimensional fields: seed metric:}
The non-zero three-dimensional scalars for the seed solution take the following form,
\be
y_1 = \frac{1}{y_2} = \frac{1}{y_3} = \sqrt{\frac{T_1}{T_2}},\label{eq:beginappendix}
\ee
\bea
\zeta^1 &=& \frac{2  \sqrt{q} \sqrt{p+q} (m \sqrt{q^2 + m^2} (p+\rho)-b q \sqrt{p^2 + m^2} x)}{T_2}, \\
\tilde \zeta_1 &=& -\frac{2 \sqrt{p}  \sqrt{p+q} (b p \sqrt{q^2 + m^2} x+m \sqrt{p^2 + m^2} (q+\rho ))}{T_1},
\eea
\bea
e^{2U} &=& \frac{m (p+q) F}{\sqrt{T_1 T_2}},
\eea
\bea
\sigma &=&
\frac{4 \sqrt{pq}}{T_1 T_2} \Big{\{}
m^2 (p-q) \sqrt{(p^2 + m^2)(q^2 + m^2)} \left(-m^2+q \rho +p (q+\rho
   )\right) \nn \\
& & + m b x (p+q) \left(m^4+\left(p^2+(q+\rho ) p+q^2+\rho ^2+q
   \rho \right) m^2-p q \rho  (p+q+\rho )\right) \nn \\
& & + b^2 x^2 (p-q) \left(m^2-p
   q\right) \sqrt{(p^2 + m^2)(q^2 + m^2)}  \nn \\
& & + m b^3 x^3 (p+q) \left(p q-m^2\right)
   \Big{\}},
\eea
where
\bea
T_1 &=& (q-p) m^3+\left(2 q p^2-b^2 x^2 p+2 (p+q) \rho  p-b^2 q x^2+(p+q) \rho ^2\right) m\nn \\ & &
+2 b p  \sqrt{(p^2 + m^2)(q^2 + m^2)} x, \\
T_2 & = & (p-q) m^3+\left(2 p q^2-b^2 x^2 q+2 (p+q) \rho  q-b^2 p x^2+(p+q) \rho ^2\right) m \nn \\ & &
-2 b q  \sqrt{(p^2 + m^2)(q^2 + m^2)} x.
\eea
\paragraph{Final 5d fields:}
Now that we have all scalars we can write the final five-dimensional solution by dualizing the appropriate scalars into one forms. For this solution we need to do only two such dualizations. We get
\be
\omega_3 = -\frac{2 b \sqrt{p} \sqrt{q} \left(1 - x^2\right) \left(m^2 (p+q+\rho )-p q \rho \right)}{m (p+q)
   \left(m^2-b^2 x^2+\rho ^2\right)} d\phi,
\ee
and
\be
A^1_{3} = \frac{2 \sqrt{p}}{\sqrt{p+q}F}\left[\sqrt{p^2 + m^2} \Delta x - \frac{b\sqrt{q^2+m^2}}{m}(\rho p - m^2)(1-x^2)\right]d\phi.
\ee
We can lift these 3d fields to 4d fields using \eqref{metric4d3d} and \eqref{vector4d3d}. The 4d vector $\check A^1_{[1]}$ are
\begin{equation}
\check A^1_{[1]} = \zeta^1(dt+\omega_3) + A_3^1  = \zeta^1 dt + ( \zeta^1  \omega_3{}_{\phi} + A_3^1{}_{\phi}) d\phi.
\end{equation}
The function $f$ required in going from 4d to 5d is $f = \left( \frac{T_2}{T_1} \right)^{\frac{1}{6}}.$ As a result the 5d metric is
\bea
ds^2_5 &=& f^{2}dz_5^2 + f^{-1} ds^2_4, \nn \\
 &=& \left( \frac{T_2}{T_1} \right)^{\frac{1}{3}}dz_5^2 + \left( \frac{T_1}{T_2} \right)^{\frac{1}{6}} \left[-\frac{m (p+q) F}{\sqrt{T_1 T_2}}(dt+\omega_\phi d\phi)^2 + \frac{\sqrt{T_1 T_2}}{m (p+q) F}ds_3^2\right], \nn \\
 &=&  T_1^{-1/3}  T_2^{1/3} dz_5^2  - T_1^{-1/3} T_2^{-2/3} (m (p+q) F)(dt+\omega_\phi d\phi)^2 \nn  \\ & &
 + \frac{T_1^{2/3}T_2^{1/3}}{m (p+q)}\left( \frac{d \rho^2}{\Delta}  + \frac{dx^2}{1-x^2}  + (1-x^2)  \frac{\Delta}{ F} d\phi^2 \right).
\eea
The 5d vectors are
\begin{align}
A^1_{[1]} &=  \zeta^1 dt + ( \zeta^1  \omega_3{}_{\phi} + A_3^1{}_{\phi}) d\phi, &
A^2_{[1]} &= 0 &
A^3_{[1]} &= 0,
\end{align}
and the 5d scalars are
\begin{align}
h_1 &= T_1^{2/3} T_2^{-2/3}, & h_2 &= T_1^{-1/3} T_2^{1/3}, & h_3  &= T_1^{-1/3} T_2^{1/3}.
\end{align}
\paragraph{6d lift and comparison with GRS:}
The 6d lift is therefore
\bea
ds^2_6 &=& e^{-\sqrt{\frac{3}{2}} \Psi} (dz_6 + A_{[1]}^1)^2 +
e^{\frac{1}{\sqrt{6}} \Psi} ds^2_5,\\
&=& \frac{T_2}{T_1}(dz_6 + A_{[1]}^1)^2 + T_1^{1/3} T_2^{-1/3}ds^2_5, \\
&=& \frac{T_2}{T_1}(dz_6 +  \zeta^1 dt + ( \zeta^1  \omega_3{}_{\phi} + A_3^1{}_{\phi}) d\phi)^2 +  dz_5^2  - \frac{(m (p+q) F)}{T_2}(dt+\omega_\phi d\phi)^2 \nn  \\ & &
 + \frac{T_1}{m (p+q)}\left( \frac{d \rho^2}{\Delta}  + \frac{dx^2}{1-x^2}  + (1-x^2)  \frac{\Delta}{ F} d\phi^2 \right),
\eea
and the three-form $F_{[3]}$ is identically zero. We are now in position to compare this metric with the GRS seed metric~\cite{Giusto:2007tt}.  The two metrics are \textit{exactly} the same. A dictionary between our notation and the notation of GRS is as follows (left-hand side is our notation and the right-hand side is GRS notation)
\bea
\frac{T_1}{m (p + q)} &=& A, \\
\frac{T_2}{m (p + q)} &=& B, \\
 \omega_3 &=& \omega^0, \\
 F &=& f_\rom{GRS}^2, \\
\zeta^1 &=& \frac{C}{B}, \\
A_3^1{} &=& \omega^1.
\eea

\paragraph{Change of gauge:}
Before we proceed to adding charges, we take care of one more
technicality at this stage.
This is a constant shift in the value of $\sigma$. This change of gauge is implemented in section 3.2 of reference~\cite{Giusto:2007tt}.
Constant shifts in $\sigma$ are implemented by the SO(4,4) generator $E_0$. We act on the previously obtained matrix $\cM$ with
\be
g_G = \exp \left[-\alpha E_0\right],
\ee
which amounts of simply shifting $\sigma$ to $\sigma + 2 \alpha$. Since $V_0$ in reference~\cite{Giusto:2007tt}
corresponds to the following combination of scalars in our approach
\be
V_0 = \frac{1}{2} ( \sigma + \zeta^1 \tilde \zeta^1),
\ee
the shift in $\sigma$ is equivalent to a shift in $V_0$,
\be
V_0 \to V_0 + \alpha.
\ee
We choose
\be
\alpha = - \sqrt{\frac{q (q^2 + m^2)}{p (p^2 + m^2)}}.
\ee

As a result the new expression for $V_0$ becomes
\be
V_0 = -\frac{1}{A}\sqrt{\frac{q (q^2 + m^2)}{p (p^2 + m^2)}}\left[ F + 2 p \left( \rho + p + \frac{q b}{m} \sqrt{\frac{p^2 + m^2}{q^2 + m^2}} x\right) \right],
\ee
and the rest of the 3d scalars remain unchanged.

For the study of asymptotic properties of the five-dimensional metric written in the main text, it is useful to first `undo' this gauge transformation and then do asymptotic expansions.

\subsubsection{Adding magnetic charges}
Now we add two further magnetic charges on the seed solution. We act on the previously obtained  matrix $\cM$ with the group element
\be
g = \exp(\alpha_2 K_{p^2}) \exp(\alpha_3 K_{p^3}),
\ee
to obtain the matrix $\cM$ for the final configuration.
To avoid notational clutter we use
\be
s_2 = \sinh \alpha_2, \qquad s_3 = \sinh \alpha_3, \qquad c_2 = \cosh \alpha_2, \qquad c_3 = \cosh \alpha_3.
\ee
To write the final configuration, we need to introduce some more notation.  For the ease of comparing with~\cite{Giusto:2007tt} we mostly use their notation in the following. Most of the quantities are already introduced above. Let us further define
\bea
D &=& B c_2^2 c_3^2 -F (c_2^2 s_3^2 + s_2^2 c_3^2) + \frac{F^2}{B}
s_2^2 s_3^2 - \frac{C^2 F}{A B}  s_2^2 s_3^2, \\
G &=& \frac{A F -C^2}{B}, \\
\tilde H_{2,3} &=& A + (A - G) s_{2,3}^2.
\eea
Using judiciously the previous notation we can write all the resulting three-dimensional scalars as follows.
\paragraph{Three-dimensional fields:}
The names $\sigma_\rom{seed}$, $\zeta^1{}_\rom{seed}$, $\tilde
\zeta^1{}_\rom{seed}$ refer to the expressions for $\sigma$, $\zeta^1$, and $\tilde
\zeta^1$ reported in the previous subsection for the seed solution. We have
\begin{align}
x_1 &= \frac{A s_2 s_3 V_0}{D + A s_2^2 s_3^2 V_0^2},  & x_2 &= c_2 s_3 \frac{C}{\tilde H_3}, & x_3 &= c_3 s_2 \frac{C}{\tilde H_2},\\
y_1 &= \frac{\sqrt{AD}}{D + A s_2^2 s_3^2 V_0^2}, &  y_2 &= \frac{\sqrt{A D}}{\tilde H_3}, &  y_3 &= \frac{\sqrt{A D}}{\tilde H_2},
\end{align}

\begin{align}
\zeta_0 &= c_2 c_3 s_2 s_3 \frac{C V_0}{D}, &
\tilde \zeta^0 &= - s_2 s_3 \frac{C F}{A D}, \\
%%%%%%%%%
\zeta_1 &= c_2 c_3 \frac{C}{D}, &
\tilde \zeta^1 &= \tilde \zeta^1_\rom{seed} + s_2^2 s_3^2 \frac{C F
V_0}{A D}, \\
%%%%%%%%%
\zeta_2 &= s_2 c_3 (B c_2^2 - F s_2^2) \frac{V_0}{D}, &
\tilde \zeta_2 &= -\frac{c_2 s_2}{ A B D} (-C^2 F s_3^2 + A (B - \tilde
F) (B c_3^2 - F s_3^2)), \\
%%%%%%%%%
\zeta_3 &= s_3 c_2 (B c_3^2 - F s_3^2) \frac{V_0}{D}, &
\tilde \zeta_3 &= -\frac{c_3 s_3}{ A B D} (-C^2 F s_2^2 + A (B - \tilde
F) (B c_2^2 - F s_2^2)),
\end{align}
and
\begin{align}
e^{2 U} &= \frac{F}{\sqrt{AD}}, &
\sigma &= \frac{A B c_2 c_3 ((B-F) (s_2^2 + s_3^2) V_0 + B
\sigma_\rom{seed})}{A (B c_2^2 - F s_2^2) (B c_3^2
- F s_3^2)-C^2 F s_2^2 s_3^2}.
\end{align}
As a next step we need to dualize pseudo-scalars  $\tilde \zeta^\Lambda$ and  $\sigma$ into 1-forms.
\paragraph{Dualizations:}
Implementing dualizations of the five scalars $\sigma$ and $\tilde \zeta^\Lambda$ is somewhat tedious. Here we give the final results:
\bea
\omega_3 &=& c_2 c_3 \omega_3^\rom{old} \\
&\equiv& c_2 c_3 \omega^0  \\
A^0_3 &=& \frac{2 b \sqrt{q} \sqrt{p+q} }{m \sqrt{m^2+p^2}
F} s_2 s_3 \left((p-q+\rho ) m^2+p q \rho \right) \left(1- x^2 \right) d\phi \\
&\equiv& s_2 s_3 \kappa^1_0 \\
A^1_3 &=& \frac{2 \sqrt{p}}{\sqrt{p+q}
   F} \left[\sqrt{m^2+p^2}  \Delta x - \frac{b\sqrt{m^2+q^2}}{m} (1-x^2) (p \rho - m^2)\right] d \phi \nn \\
   &\equiv& \omega^1 \\
A^2_3 &=&  \frac{2}{F} q s_2 c_2 \left[ \Delta x +
\sqrt{\frac{q^2 + m^2}{p^2 + m^2}} \frac{b}{m} (p \rho - m^2)(1-x^2)\right] \nn \\
   &\equiv& -s_2 c_2 \kappa^{0}_{0} \\
A^3_3 &=& -s_3 c_3 \kappa^{0}_{0}
\eea
where $\omega^0$, $\kappa^1_0$, $\omega^1$, and $\kappa^{0}_{0}$  are the same as in equations (2.49), (3.51), (2.50), and (3.52) respectively of~\cite{Giusto:2007tt}, and $\omega_3^{\rom{old}}$ is the same as
 $\omega_3$ of the solution obtained in the previous subsection.
Thus we have successfully obtained all three-dimensional
fields to describe it as a solution of 5d U(1)$^3$ supergravity.
 In section \ref{summary} we listed all the five-dimensional fields.

\section{\texorpdfstring{SO(4,4)$/$(SO(2,2)$\times$}{}
SO(2,2)) coset in three dimensions}
\label{GroupTheory}

\subsection{Parametrization}
\label{appSO44}
The metric $G_{ab}$ in equation \eqref{metric3dcoset} in our conventions
 is
\bea
 G_{ab}d\varphi^a d\varphi^b &=& 4 dU^2 + 4 g_{I \bar{J}}dz^I dz^{\bar J} +
 \frac{1}{4}e^{-4U} \left( d\sigma +  \tilde \zeta_\Lambda d \zeta^\Lambda -
 \zeta^\Lambda d \tilde \zeta_\Lambda \right)^2 \label{cmap} \\
 && \hspace{-2cm} + e^{-2U}\left[ (\mbox{Im} N)_{\Lambda \Sigma}d\zeta^\Lambda
 d\zeta^\Sigma + ((\mbox{Im} N)^{-1})^{\Lambda \Sigma} \left( d\tilde
 \zeta_\Lambda +(\mbox{Re} N)_{\Lambda \Xi} d\zeta^\Xi \right)  \left( d\tilde
 \zeta_\Sigma +(\mbox{Re} N)_{\Sigma \Xi} d\zeta^\Xi \right) \nonumber \right].
\eea
This metric can  be parameterized in the Iwasawa gauge by a coset
element $\cV$ of SO(4,4)/(SO(2,2) $\times$
SO(2,2)) as ~\cite{Bossard:2009we}
\be
\label{iwa}
\cV = e^{- U \, H_0} \cdot \left(
\prod_{I=1,2,3}
e^{-\frac{1}{2} (\log y^I) H_I} \cdot e^{ - x^I E_I} \right) \cdot
e^{-\zeta^\Lambda E_{q_\Lambda}-  \tilde \zeta_\Lambda E_{p^\Lambda}}\cdot
e^{-\frac{1}{2}\sigma E_0}.
 \ee
To see this, note that metric \eqref{cmap} can also be obtained as
\be
G_{ab} d\varphi^a d\varphi^b = \mathrm{Tr}( P_*\; P_* )
\ee
where
\be
P_* = \frac12 ( \theta + \eta'\, \theta^T {\eta'}^{-1})\ ,\quad
\eta' = {\rm diag}(-1,-1,1,1,-1,-1,1,1).
 \label{dsp}
\ee
Here $\eta'$ is the quadratic form preserved by SO(2,2)$\times$SO(2,2) (in our basis), and
$\theta= d\cV \cdot \cV^{-1}$ is the Mauer-Cartan 1-form.

Next we define the matrix $\cM$ as
$\cM = (\cV^\sharp) \cV$. The operation $\sharp$ refers to the generalized transposition,
\be
 \theta^\sharp = \eta' \theta^T {\eta'}^{-1},
\ee
for any $\theta \in \mf{so}(4,4)$.

Now we list an explicit matrix representation of $\mf{so(4,4)}$. Denoting $E_{ij}$ the $8 \times
8$ matrix with 1 in the position $(i,j)$ and 0 elsewhere, the Lie algebra generators are
\be
\begin{array}{|c|c|}
\hline
H_0 = E_{33} + E_{44} - E_{77} - E_{88} & H_1 = E_{33} - E_{44} - E_{77} + E_{88}\nn \\
H_2 = E_{11} + E_{22} - E_{55} - E_{66} & H_3 = E_{11} - E_{22} - E_{55} + E_{66} \\ \hline
%\eea
%\bea
E_0 = E_{47} - E_{38} & E_1 = E_{87} - E_{34}\nn \\
E_2 = E_{25} - E_{16} & E_3 = E_{65}-E_{12}  \\ \hline
F_0 = E_{74} - E_{83} &  F_1 = E_{78} - E_{43}\nn \\
F_2 = E_{52} - E_{61} & F_3 = E_{56} - E_{21}\\ \hline
E_{q{}_0} = E_{41} - E_{58} & E_{q{}_1} = E_{57} - E_{31} \nn \\
E_{q{}_2} = E_{46} - E_{28} & E_{q{}_3} = E_{42} - E_{68}   \\ \hline
F_{q{}_0} = E_{14} - E_{85} & F_{q{}_1} = E_{75} - E_{13} \nn \\
F_{q{}_2} = E_{64} - E_{82} & F_{q{}_3} = E_{24} - E_{86}      \\ \hline
E_{p{}^0} = E_{17} - E_{35} & E_{p{}^1} = E_{18} - E_{45}\nn  \\
E_{p{}^2} = E_{67} - E_{32} & E_{p{}^3} = E_{27} - E_{36}      \\ \hline
F_{p{}^0} = E_{71} - E_{53} & F_{p{}^1} = E_{81} - E_{54} \nn  \\
F_{p{}^2} = E_{76} - E_{23} & F_{p{}^3} = E_{72} - E_{63} \\ \hline
\end{array}
\ee
This basis is identical to the one given in~\cite{Virmani:2012kw, Bossard:2009we}.

\subsection{\texorpdfstring{$\mathrm{SO}(2,2)\times \mathrm{SO}(2,2)$}{} generators}
\label{appSO22}
The involution of Lie algebra that defines the coset is simply
\be
\tau(\theta) := - \theta^\sharp.
\ee
Lie algebra generators belonging to the denominator subgroup $\mathrm{SO}(2,2)\times \mathrm{SO}(2,2)$ are those which are invariant under the involution; their explicit form is $x + \tau(x)$. Such generators are
\be
\begin{array}{|c|c|}
\hline
K_0 = E_0 - F_0 & K_1 = E_1 - F_1 \\
K_2 = E_2 - F_2 & K_3 = E_3 - F_3 \\ \hline
K_{q{}_0} = E_{q{}_0} + F_{q{}_0} & K_{q{}_1} = E_{q{}_1} +F_{q{}_1} \\
K_{q{}_2} = E_{q{}_2} +F_{q{}_2} & K_{q{}_3} = E_{q{}_3} +F_{q{}_3} \\ \hline
K_{p{}^0} = E_{p{}^0} + F_{p{}^0} & K_{p{}^1} = E_{p{}^1} +F_{p{}^1} \\
K_{p{}^2} = E_{p{}^2} +F_{p{}^2} & K_{p{}^3} = E_{p{}^3} +F_{p{}^3} \\ \hline
\end{array}
\ee
To see that these generators form an
$\mathfrak{so}(2,2)\oplus \mathfrak{so}(2,2) \equiv [\mathfrak{sl}(2, \RR)^4]$ Lie algebra, consider the following combinations of the above generators~\cite{Katsimpouri:2013wka}

\be
\begin{array}{|c|}
\hline
h_1 = \frac{1}{2}(-K_{q{}_0} + K_{p{}^1} + K_{p{}^2} + K_{p{}^3}) \\
h_2 = \frac{1}{2}(+K_{q{}_0} - K_{p{}^1} + K_{p{}^2} + K_{p{}^3}) \\
h_3 = \frac{1}{2}(+K_{q{}_0} + K_{p{}^1} - K_{p{}^2} + K_{p{}^3}) \\
h_4 = \frac{1}{2}(+K_{q{}_0} + K_{p{}^1} + K_{p{}^2} - K_{p{}^3}) \\ \hline
e_1 = \frac{1}{4}(-K_0 + K_1 + K_2 + K_3 + K_{q{}_1} + K_{q{}_2} + K_{q{}_3} + K_{p{}^0}) \\
e_2 = \frac{1}{4}(+K_0 - K_1 + K_2 + K_3 + K_{q{}_1} - K_{q{}_2} - K_{q{}_3} + K_{p{}^0})\\
e_3 = \frac{1}{4}(+K_0 + K_1 - K_2 + K_3 - K_{q{}_1} + K_{q{}_2} - K_{q{}_3} + K_{p{}^0})\\
e_4 = \frac{1}{4}(+K_0 + K_1 + K_2 - K_3 - K_{q{}_1} - K_{q{}_2} + K_{q{}_3} + K_{p{}^0}) \\ \hline
f_1 = \frac{1}{4}(+K_0 - K_1 - K_2 - K_3 + K_{q{}_1} + K_{q{}_2} + K_{q{}_3} + K_{p{}^0}) \\
f_2 = \frac{1}{4}(-K_0 + K_1 - K_2 - K_3 + K_{q{}_1} - K_{q{}_2} - K_{q{}_3} + K_{p{}^0})\\
f_3 = \frac{1}{4}(-K_0 - K_1 + K_2 - K_3 - K_{q{}_1} + K_{q{}_2} - K_{q{}_3} + K_{p{}^0})\\
f_4 = \frac{1}{4}(-K_0 - K_1 - K_2 + K_3 - K_{q{}_1} - K_{q{}_2} + K_{q{}_3} + K_{p{}^0}) \\ \hline
\end{array}
\ee
These linear combinations manifest the four commuting $\mathfrak{sl}(2, \RR)$s --- $(h_i, e_i, f_i)$ with commutation relations $[h_i, e_i] = 2 e_i, [h_i, f_i] = -2 f_i, [e_i, f_i] =  h_i$.

\bibliographystyle{toine}
\bibliography{Papers}

\end{document}